\documentclass[11pt,a4paper]{article}
\pdfoutput=1   

\usepackage{alpha}
\hypersetup{%
   pdftitle = {},
   pdfauthor = {Mattia Dalla Brida, Patrick Fritzsch, Tomasz Korzec, Alberto Ramos, Stefan Sint, Rainer Sommer},
   pdfkeywords = {Lattice QCD, Renormalization group}
}%
\usepackage{macros_alpha}
\usepackage{caption}
\usepackage{subcaption}
\usepackage[section]{placeins}

\begin{document}

%
%
\preprintno{%
CERN-TH-2018-060\\
DESY 18-044\\
WUB/18-01\\ 
}
%
%
\title{A non-perturbative exploration \\ of the high energy regime in $\Nf=3$ QCD}
%
%
\collaboration{\includegraphics[width=2.8cm]{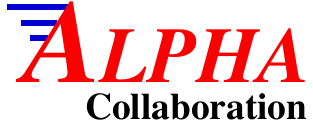}}
%
%
\author[milano]{Mattia~Dalla~Brida}
\author[cern]{Patrick~Fritzsch} %
\author[wup]{Tomasz~Korzec} %
\author[dublin]{\\Alberto~Ramos} 
\author[dublin]{Stefan~Sint}%
\author[desy,hu]{Rainer~Sommer}
%
%
\address[milano]{Dipartimento di Fisica, Universit\`a di Milano-Bicocca
                    and INFN, sezione di Milano-Bicocca, \\ Piazza della Scienza 3,
                    I-20126 Milano, Italy}
\address[cern]{CERN, Theoretical Physics Department, CH-1211 Geneva 23, Switzerland}
\address[wup]{Department of Physics, Bergische Universit\"at Wuppertal, Gau\ss str. 20,
D-42119 Wuppertal, Germany}
\address[dublin]{School of Mathematics \& Hamilton Mathematics Institute, Trinity College Dublin, Dublin
  2, Ireland}
\address[desy]{John von Neumann Institute for Computing (NIC), DESY, Platanenallee~6, 15738 Zeuthen, Germany}
\address[hu]{Institut~f\"ur~Physik, Humboldt-Universit\"at~zu~Berlin, Newtonstr.~15, 12489~Berlin, Germany}
%

%
\begin{abstract}
Using continuum extrapolated lattice data we trace a family of running couplings
in three-flavour QCD over a large range of scales from about 4 to 128 GeV.
The scale is set by the finite space time volume so that recursive finite
size techniques can be applied, and Schr\"odinger functional (SF)
boundary conditions enable direct simulations in the chiral limit.
Compared to earlier studies we have improved on both statistical and systematic errors.
Using the SF coupling to implicitly define a reference scale $1/L_0\approx 4$~GeV 
through $\bar{g}^2(L_0) =2.012$, we quote $L_0 \Lambda^{\Nf=3}_{\MSbar} =0.0791(21)$.
This error is dominated by statistics; in particular, the 
remnant perturbative uncertainty is negligible and very well controlled, 
by connecting to infinite renormalization scale 
from different scales $2^n/L_0$ for $n=0,1,\ldots,5$.
An intermediate step in this connection may involve any member of a one-parameter family of SF couplings.
This provides an excellent opportunity for tests of perturbation theory some of which 
have been published in a letter~\cite{Brida:2016flw}.
The results indicate that for our target precision of 3 per cent in 
$L_0 \Lambda^{\Nf=3}_{\MSbar}$, a reliable estimate of the truncation
error requires non-perturbative data for a sufficiently large range of
values of $\alpha_s=\bar{g}^2/(4\pi)$. In the present work we reach
this precision by studying scales that vary by a factor $2^5= 32$, 
reaching down to $\alpha_s\approx 0.1$. We here provide the details of our analysis 
and an extended discussion.\\
\end{abstract}
%
%
%
%


\maketitle

\tableofcontents

\newpage


\section{Introduction}

The Standard Model seems to describe all high energy physics
experiments carried out to date, in some cases with extraordinary
accuracy~(cf.~\cite{Patrignani:2016xqp} for the most recent PDG
review). For processes involving the strong interactions the precision
is usually less impressive, due to our limited ability to extract
quantitative information from QCD.  One of the main tools is
perturbation theory (PT) in the strong coupling, $\alpha_s$, and there
has been significant progress in high order perturbative QCD
calculations, with renormalization group functions now available up to
5-loop order in the $\MSbar$-scheme~\cite{Chetyrkin:2017bjc,Baikov:2016tgj,Luthe:2017ttc,MS:4loop1,Czakon:2004bu}.
However, before a perturbative result can be confronted with
experimental observables, the transition from quarks and gluons to
hadronic degrees of freedom needs to be modelled in some way. Such
models come in various shapes and forms, from
``hadronisation Monte Carlo'' in jet physics to ``quark hadron
duality'' in QCD sum rules.  A common problem then consists in
assigning systematic errors to the model assumptions. A further issue
is the reliability of PT itself, given that the series is only
asymptotic. To some extent, the reliability can be assessed within PT
itself, by comparing different orders of the expansion, or by
increasing the energy scale, $\mu$, such that $\alpha_s(\mu)$ becomes
small, due to asymptotic freedom. Unfortunately, the rapidly
increasing complexity of higher order calculations means that
typically only a few terms in the perturbative series are available.
In addition, the energy scale is often defined by the kinematics of
the physical process under consideration. The variation of the scale
is then rather limited and to assign an error to the perturbative
result is difficult\footnote{For a recent discussion in the context of
$\alpha_s$-determinations cf.~ref.~\cite{Salam:2017qdl}.}.  

In this work we carry out a systematic investigation into the reliability of PT. 
We do this by directly comparing non-perturbative QCD observables to their perturbative expansions,
over a wide range of scales.  Lattice QCD, together with a careful
treatment of the continuum limit, is currently the only way to obtain
such non-perturbative results, subject only to standard assumptions
such as locality and universality. The main reason why this is rarely done
is the usual limitation of any numerical approach: on a finite system
it is very expensive to simultaneously resolve very different length scales. 
Most lattice QCD projects aim at hadronic low energy physics, and 
the space-time volume, $L^4$, must then measure several femto metres across in order 
not to distort the hadronic states of interest.
At least for massive single particle states, the finite volume effects are exponentially
suppressed \cite{Luscher:1985dn} and one may then pretend to be in infinite space time volume, up to a systematic
error which is often below the percent level.
On the other hand, with current lattice resolutions of, say, $L/a < 100$ 
this means that the cutoff scale set by the inverse lattice spacing, $1/a$, reaches a few GeV at most,
and the deep perturbative high energy regime seems out of reach.
It is important to realize that this limitation is only due to the requirement 
that the lattice covers a physically large space-time volume.
If this constraint is lifted, there is nothing that prevents 
simulations at very high energies, albeit in physically tiny space-time volumes.
The observables\footnote{Here by observable we mean some finite quantity
defined by the Euclidean path integral, that can be estimated in a Monte Carlo simulation
of lattice QCD.} we consider in this situation are all normalized as effective
couplings, which run with $L$, the scale set by the finite space-time volume. In order
to achieve this we set all quark masses to zero and scale all other dimensionful parameters
proportionally to $L$, thereby obtaining a mass-independent scheme. In the high-energy
regime, PT can be used to relate to more commonly used schemes
such as the $\MSbar$ scheme of dimensional regularization.
Moreover, by combining the idea of a finite volume  scheme for the coupling
with recursive step-scaling techniques~\cite{Luscher:1991wu,Jansen:1995ck}, one
may both determine the scale $L$ in units of some hadronic scale, 
and reach the perturbative high energy regime without ever requiring enormous lattice resolutions, $L/a$.
Obviously, the finite space-time volume constitutes an integral part 
in the definition of these observables. PT must then be adapted to this situation.
While the Euclidean space-time signature used in lattice QCD is advantageous
in PT, all the sophisticated tools of standard PT in (infinitely extended) momentum space 
are of limited use. 

As part of the project to determine $\alpha_s(m_Z)$ from low energy hadronic input in
3-flavour QCD~\cite{Bruno:2017gxd,DallaBrida:2016kgh,Brida:2016flw},
our collaboration has applied these techniques to a 1-parameter family
of finite volume couplings in Schr\"odinger functional (SF)
schemes, for which the 3-loop $\beta$-function is
known~\cite{Bode:1998hd,Bode:1999sm,Christou:1998wk,Christou:1998ws}.
We have measured these couplings in numerical simulations and for a
range of lattice sizes with unprecedented precision. Extrapolation to
the continuum limit of this data allows us to carry out stringent
tests of renormalized perturbation theory for energy scales ranging
from about $4\,\GeV$ to $128\,\GeV$.
A first account of our results has appeared in a letter \cite{Brida:2016flw} 
and we here provide the details of this work and a more extended analysis.

\begin{figure}
  \centering
  \includegraphics[width=0.85\textwidth]{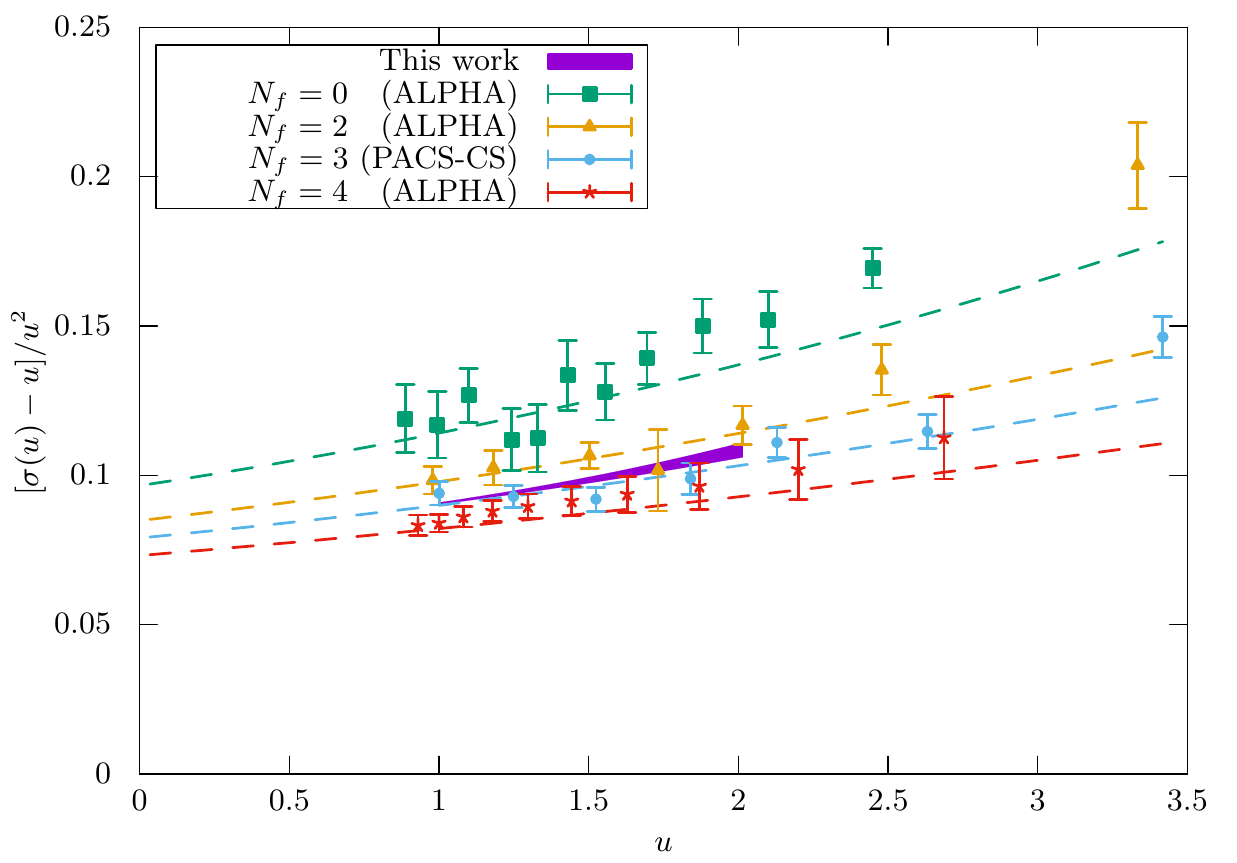}
  \caption{The step scaling function $\sigma(u)$, a discrete
      version of the $\beta$-function, defined
  in \eq{eq:sigdef}. The combination shown here yields
  directly the lowest order cofficient, $b_0$ 
  of the $\beta$-function
  as $(\sigma(u)-u)/u^2 = 2 b_0\ln\,2 +\rmO(u)$.
  The dashed lines show the perturbative 2-loop behavior. 
  The purple 1-sigma band shows our result (fit~C in table~\ref{tab:Sigfits}).
  Data points for $\nf=0,2,3,4$ are taken from the literature
  \cite{Capitani:1998mq,DellaMorte:2004bc,Aoki:2009tf,Tekin:2010mm}.
  }
  \label{fig:all_nf}
\end{figure}

The technique, used earlier for between 
$\nf=0$ and $\nf=4$ quark flavours \cite{Capitani:1998mq,DellaMorte:2004bc,Aoki:2009tf,Tekin:2010mm},  
allows one to non-perturbatively 
verify the close-to perturbative running of the coupling 
and observe the small effects of dynamical quarks, as illustrated 
in \fig{fig:all_nf}. 
A preview of our final result is included 
in the figure, demonstrating our advanced
precision.

The paper is organized as follows: section~2 uses a continuum language
to explain how our QCD observables are defined and collects the
relevant perturbative results from the literature. We also comment on
``non-perturbative effects'' which are associated with
secondary minima of 
the action. Section~3 then presents the lattice set-up, the numerical
simulations and statistics produced, and discusses the perturbative
improvement of the data.  The impatient reader might skip this section
and directly pass to section 4.  There, after the discussion of the
continuum extrapolated results and associated systematic errors, the
comparison to renormalized perturbation theory is performed before we
conclude in Section~5. Finally, a technical appendix presents the models we used for the
sensitivity of the data to a variation of the two O($a$) boundary
counterterm coefficients $\ct$ and $\ctt$.

\section{SF couplings}
\label{sec:coupling}

In order to apply the recursive step-scaling techniques to lattice QCD, it is
desirable to define renormalized QCD couplings in a finite space-time
volume, $L^4$, and in the chiral limit. Such finite volume 
renormalization schemes are quark mass independent 
by construction~\cite{Weinberg:1951ss}, and the renormalization 
scale is set by $\mu=1/L$. 
It is then possible to apply recursive finite size
scaling methods and trace the scale evolution over a wide range
without the need for very large lattice sizes,
$L/a$~\cite{Luscher:1991wu}.  Still, these requirements leave many
options, such as the boundary conditions for the fields and the exact
choice of observable. We here choose Schr\"odinger functional boundary
conditions~\cite{Luscher:1992an,Sint:1993un}: these introduce a gap in
the spectrum of the Dirac operator, so that numerical simulations can
be performed directly at zero quark masses, without the need for
any chiral extrapolation. Moreover, perturbation theory 
remains tractable in this framework, as the absolute minimum of the
action is unique up to gauge equivalence.  For the observable we
choose the traditional SF coupling~\cite{Luscher:1993gh,Sint:1995ch}
and a 1-parameter family of close relatives~\cite{Sint:2012ae}. 
The most important reason for this choice is the existence of a 2-loop
calculation in this case~\cite{Bode:1998hd,Bode:1999sm}, which, in combination
with~\cite{Christou:1998wk,Christou:1998ws} allows to infer the 3-loop
$\beta$-function for these schemes. Furthermore, the values
of the 3-loop $\beta$-function coefficients are reasonable and
enable us to make contact with the asymptotic perturbative regime
at energy scales in the range O(10--100) GeV.

In the future one might also consider the more recent coupling 
definitions based on the gradient
flow~\cite{Luscher:2010iy,Fritzsch:2013je}.  
The QCD 3-loop $\beta$-function is currently
known in the case of infinite space-time volume~\cite{Harlander:2016vzb}, 
and there is progress for the case of a finite volume with SF boundary
conditions~\cite{Fritzsch:2013je} using numerical stochastic perturbation
theory~\cite{Brida:2013mva,DallaBrida:2016dai, DallaBrida:2017tru}.
These results seem to point to a 3-loop $\beta$-function
coefficient which is significantly larger than in the
$\MSbar$- and SF-schemes. This indicates that gradient flow couplings 
may not be ideal for matching with the asymptotic perturbative regime.
Furthermore, cutoff effects are typically larger with 
the GF couplings than with the traditional SF
coupling~\cite{DallaBrida:2016kgh},  
so that larger lattice sizes are required. This partially offsets 
other computational advantages. Obviously, further studies are required and 
one should re-assess the situation once more perturbative
information becomes available.

\subsection{SF$_\nu$ schemes}

In the continuum the Schr\"odinger functional is defined as the
Euclidean path integral,
\begin{equation} {\cal Z}[C,C']= \int D[\Lambda] \int
D[A,\psi,\psibar]{\rm e}^{-S[A,\psi,\psibar]},
   \label{eq:SF_functional_integral}
\end{equation} with the Euclidean continuum action $S=S_g+S_f$,
\begin{eqnarray}
  \label{eq:S_g} S_g &=& -\frac{1}{2g_0^2} \int_0^L\rmd x_0
\int_0^L\rmd^3{\bfx} \tr\{F_{\mu\nu}(x) F_{\mu\nu}(x)\}\,,\\ S_f &=&
\int_0^L\rmd x_0 \int_0^L\rmd^3{\bfx}\, \psibar(x)(\gamma_\mu
D_\mu+m)\psi(x)\,.
\end{eqnarray} Here, $g_0$ denotes the bare coupling constant,
$F_{\mu\nu}$ is the field tensor associated with the gauge field
$A_\mu$,
\begin{equation} F_{\mu\nu}=\partial_\mu A_\nu-\partial_\nu A_\mu
+[A_\mu,A_\nu],
\end{equation} and $D_\mu=\partial_\mu+A_\mu+i\theta_\mu/L$ is the
covariant derivative acting on the quark fields. It includes a
constant U(1) background field which we set to $\theta_\mu= (1-\delta_{\mu
0})\theta$, with the choice $\theta=\pi/5$.  In the spatial directions $L$-periodic
boundary conditions are imposed on all fields. At the time boundaries 
the fermionic fields satisfy~\cite{Sint:1993un}
\begin{equation} P_+\psi\vert_{x_0=0}=0=P_-\psi\vert_{x_0=L},\qquad
\psibar P_-\vert_{x_0=0} =0= \psibar P_+\vert_{x_0=L},
\end{equation} with the projectors $P_\pm =\frac12(1\pm\gamma_0)$.
For the gauge field one has
\begin{equation} A_k|_{x_0=0} = C^{\Lambda}_k,\qquad A_k|_{x_0=L} =
C'_k,\qquad k=1,2,3,
   \label{eq:bcs_gaugefields}
\end{equation} with the boundary values $C_k$ and $C_k'$.  The
boundary condition at $x_0=0$ refers to the gauge transformed field,
\begin{equation} C_k^{\Lambda}(\bfx) = \Lambda(\bfx) C_k(\bfx)
\Lambda(\bfx)^\dagger +
\Lambda(\bfx)\partial_k\Lambda(\bfx)^\dagger\,.
\end{equation} The integration over the SU(3)-valued and spatially
periodic gauge functions $\Lambda(\bfx)$ in
Eq.~(\ref{eq:SF_functional_integral}) ensures gauge invariance of the
Schr\"odinger functional.  The spatially periodic $\Lambda(\bfx)$ fall
into different topological sectors labelled by an integer $n$,
\begin{equation} n = \frac{1}{24\pi^2} \int_0^L \dd^3{\bfx}\,
\epsilon_{ijk} \tr\!\left\{(\Lambda \partial_i\Lambda^{-1}) (\Lambda
\partial_j\Lambda^{-1})(\Lambda \partial_k\Lambda^{-1})\right\},
\end{equation} which is related to the topological charge of the gauge
field,
\begin{equation} Q[A] = -\dfrac{\epsilon_{\mu\nu\rho\sigma}}{32\pi^2}
\int {\rm d}^4x\, {\rm tr}\{F_{\mu\nu}(x) F_{\rho\sigma}(x)\}\,,
\end{equation} through $n = - Q[A]$, provided the Chern-Simons action
of the boundary gauge fields $C_k^{}$, $C'_k$ vanishes (which is the
case for the choice below). The value of the gauge action in each
sector $n$ is then subject to the usual instanton
bound~\cite{Luscher:1992an}
\begin{equation} 
  g_0^2 S_g[A] \ge 8\pi^2|Q[A]|\,.
 \label{eq:instbd}
\end{equation} 
Using the gauge invariance of the Schr\"odinger
functional under the transformations,
\begin{eqnarray} 
A_\mu(x) &\rightarrow& \Omega(x) A_\mu
\Omega(x)^\dagger + \Omega(x) \partial_\mu \Omega(x)^\dagger,\\
\Lambda(x) &\rightarrow& \Omega(0,\bfx) \Lambda(\bfx)\,,\\ \psi(x)
&\rightarrow& \Omega(x) \psi(x)\,,\\ \psibar(x) &\rightarrow&
\psibar(x)\Omega(x)^\dagger\,,
\end{eqnarray} 
one may convert the integral over gauge functions
$\Lambda$ to a sum over $n$, with $\Lambda$ in
Eq.~(\ref{eq:bcs_gaugefields}) replaced by fixed representatives
$\Lambda_n$ for each topological sector. In particular one often sets
$\Lambda_0 = 1$.

We now focus on Abelian and spatially constant boundary gauge
fields,
\begin{equation} C^{}_k(\bfx)=\frac{i}{L}\phi, \qquad C'_k(\bfx)
=\frac{i}{L}\phi', \qquad k=1,2,3,
 \label{eq:bc_phases}
\end{equation} with traceless and diagonal $3\times 3$-matrices $\phi$
and $\phi'$. Their diagonal elements
\begin{xalignat}{2} \phi^{}_1 & = \eta-\frac{\pi}{3}, & 
&\phi'_1 = -\eta-\pi, \nonumber\\ \phi^{}_2 & =
\eta\left(\nu-\frac12\right), & 
& \phi'_2 =
\eta\left(\nu+\frac12\right)+\frac{\pi}{3}, \label{eq:phi_phiprime}\\
\phi^{}_3 & = -\eta\left(\nu+\frac12\right)+\frac{\pi}{3}, & 
&\phi'_3 = -\eta\left(\nu-\frac12\right)+\frac{2\pi}{3}, \nonumber
\end{xalignat} still depend on 2 real parameters, $\eta$ and $\nu$.
In the temporal gauge and the topological charge zero sector the field
equations with these boundary conditions are solved by,
\begin{equation} B_0=0,\qquad B_k= C_k +
\frac{x_0}{L}\left(C_k'-C_k\right), \qquad k=1,2,3,
  \label{eq:contBF}
\end{equation} which corresponds to a constant chromo-electric field,
\begin{equation} G_{0k}=\partial_0 B_k = \frac{C_k'-C_k}{L}=
\frac{i(\phi'-\phi)}{L^2}, \qquad k=1,2,3.
\end{equation} Inserting the field tensor into the gauge action,
$S_g$, one obtains
\begin{equation} S_g[B]=
\frac{3}{g_0^2}\sum_{\alpha=1}^3(\phi'_\alpha-\phi^{}_\alpha)^2 =
\frac{18}{g_0^2}\left(\eta+\frac{\pi}{3}\right)^2,
\label{eq:Sclass}
\end{equation} which, for given $\eta$ (and independently of $\nu$)
constitutes the absolute minimum of the action~\cite{Luscher:1992an}.
One may thus define the effective action as a function of this
background field,
\begin{equation} \Gamma[B]=-\ln {\cal Z}[C',C],
\end{equation} 
and its perturbative expansion,
\begin{equation} \Gamma[B] \, \buildrel g_0\rightarrow0\over\sim\,\,
\frac{1}{g_0^2}\Gamma_0[B]+ \Gamma_1[B]+\rmO(g_0^2),
\label{eq:coup_exp}
\end{equation} 
with $\Gamma_0[B]=g_0^2 S_g[B]$.  The SF couplings
$\gbar^2_\nu(L)$ can be defined through
\begin{equation} {\frac{\partial\Gamma[B]}{\partial\eta}}
\biggl\vert_{\eta=0}= \frac{k}{\gbar^2_\nu(L)}\,, \qquad
k={\frac{\partial\Gamma_0[B]}{\partial\eta}}
\biggl\vert_{\eta=0}=12\pi\,.
  \label{eq:coupling}
\end{equation} 
In fact the $\nu$-dependence is explicit,
\begin{equation} \frac{1}{{\bar g}_\nu^2(L)} = \frac{1}{{\bar
g}^2(L)}-\nu \bar v(L),
\end{equation} since both $1/\gbar^2(L)$ and $\bar{v}(L)$ are
$\nu$-independent.  In terms of the effective action, $\bar{v}(L)$
reads
\begin{equation} \bar{v}(L) = -\frac{1}{k} \left.\frac{\partial^2
\Gamma[B]}{\partial\nu\partial\eta}\right\vert_{\eta=\nu=0}\,.
 \label{eq:v_definition}
\end{equation} Note that the $\nu$-independence of $\Gamma_0[B]$,
implies that $\bar{v}(L)$ has a perturbative expansion starting at
O($1$).  This ensures the correct normalization of the whole
1-parameter family of couplings, namely $\gbar_\nu^2=g_0^2$ to lowest
order. Finally we remark that the entire 1-parameter family is
determined by the expectation values,
\begin{equation} \frac{k}{\bar{g}^2} = \left\langle
\left.\frac{\partial S}{\partial\eta}\right\vert_{\eta=\nu=0}
\right\rangle,\qquad \bar{v} = \frac{-1}{k}\left\langle
\left.\frac{\partial^2 S}{\partial\nu
\partial\eta}\right\vert_{\eta=\nu=0} \right\rangle\,,
\label{eq:expectval}
\end{equation} defined in terms of the functional integral,
Eq.~(\ref{eq:SF_functional_integral}), at $\nu=0$. Once the lattice
regularization is in place both quantities will thus become
observables in numerical simulations. 

\subsection{$\beta$-functions and perturbative relations to the
$\MSbar$-coupling}

The SF couplings are defined independently of perturbation theory and
thus the same is true for their $\beta$-functions,
\begin{equation} \beta(\gbar_\nu) = - L
\frac{\partial\gbar_\nu}{\partial L} 
\buildrel {\gbar_\nu}\rightarrow0\over\sim\, 
-\gbar_\nu^3\sum_{k=0}^\infty b_k
{\gbar_\nu}^{2k}\,,
\end{equation} where the asymptotic expansion on the r.h.s.~starts out
with the standard universal coefficients $b_{0,1}$ for $\Nf=3$ QCD,
  \begin{equation}
    (4\pi)b_0= 9/(4\pi), \qquad (4\pi)^2b_1 =
4/\pi^2,
\end{equation} and the 3-loop coefficient is given by
\begin{eqnarray}
  \label{eq:b2nu}
  (4\pi)^3b^\nu_{2} &=& -\left(0.064(27) + \nu \times
                        1.259(10)\right)\,.
\end{eqnarray}
The 3-loop coefficient has been obtained by matching
the coupling at the 2-loop level to the $\MSbar$-scheme, where the
$\beta$-function is now even known to 5-loop order ($b_3$ and $b_4$ in
our notation)~\cite{Chetyrkin:2017bjc,Baikov:2016tgj,Luthe:2017ttc,MS:4loop1,Czakon:2004bu}.
For later use we collect the numerical values for $\Nf=3$ QCD in table~\ref{tab:bcoef},
together with the SF$_\nu$ scheme results for the 3 choices of the parameter,  $\nu=-0.5,0,0.3$,
which we selected for more detailed analysis in Section~\ref{sec:cont}.
\begin{table}
  \centering
  \begin{tabular}{llll}
    \toprule
    Scheme& $(4\pi)^3b_2$& $(4\pi)^4b_3$& $(4\pi)^5b_4$ \\
    \midrule
    SF ($\nu=-0.5$)  &$\phantom{-}0.5655$ &-- &--\\
    SF ($\nu=0$)     &$-0.064$            &-- &--\\ 
    SF ($\nu=0.3$)   &$-0.4417$           &-- &--\\
    \midrule
    $\overline{\rm MS}$ & 0.324447 & 0.484842& 0.416059\\
    \bottomrule
  \end{tabular}
  \caption{Coefficients in the asymptotic expansion of the
    $\beta$-function in different schemes. Note that the universal
    coefficients for $\Nf=3$ are $(4\pi)b_0 \approx 0.716197$,
    $(4\pi)^2b_1\approx0.405285$.}
  \label{tab:bcoef}
\end{table}
Comparing the $\MSbar$ to the SF$_\nu$ scheme we note
that the respective 3-loop $\beta$-functions coincide for $\nu \approx -0.3$.  
In general, $\nu$-values of O($1$) are reasonable from a
perturbative point of view.

Closely related to the $\beta$-functions are the step-scaling
functions which connect couplings at scales which differ by a factor 2. 
Defining
\begin{equation} 
   \sigma(u) = \gbar^2(2L)\vert_{u=\gbar^2(L),m(L)=0},
   \label{eq:sigdef}
\end{equation} 
the precise relationship is,
\begin{equation} \int_{\sqrt{u}}^{\sqrt{\sigma(u)}}\frac{{\rm d}g}{\beta(g)} = -\ln 2\,,
   \label{eq:beta_sigma}
\end{equation} and the perturbative expansion of $\sigma(u)$,
\begin{equation} \sigma(u) = u + s_0 u^2 + s_1 u^3 + s_2 u^4 +
\ldots\,,
   \label{eq:sigcontpert}
\end{equation} is thus determined in terms of the coefficients of the
$\beta$-function, with the first 3 given by
\begin{equation} 
s_0 = 2b_0\ln 2,\quad s_1 = s_0^2 + 2b_1 \ln 2,\quad
s_2 = s_0^3 + 10 b_0 b_1 (\ln 2)^2 +2b_2 \ln 2\,.
  \label{eq:s012}
\end{equation} 
Finally, we quote the relation between the SF and the
$\MSbar$ couplings, in terms of $\alpha = \gbar^2/(4\pi)$ at the
scales $\mu=1/L$ and $s\mu$, respectively, with $s>0$. One finds
\begin{equation} 
\alpha_\MSbar(s\mu) = \alpha(\mu) + c_1(s)
\alpha^2(\mu) + c_2(s) \alpha^3(\mu) + \ldots
\end{equation} 
with (for $\Nf=3$)~\cite{Luscher:1992an,Sint:1995ch,Bode:1998hd,Bode:1999sm,Christou:1998wk,Christou:1998ws}
\begin{eqnarray} c_1(s) &=& - 8\pi b_0 \ln(s) + 1.3752097(26)\,,\\
c_2(s)-\left(c_1(s)\right)^2 &=&  - 32\pi^2 b_1 \ln(s) + 1.320(30) \,.
\end{eqnarray} In order to connect to the SF$_\nu$ couplings for
$\nu\ne 0$ we need the expansion of $\bar{v}$ in the coupling
$\bar{g}$. Defining
\begin{equation} \omega(u) = \bar{v}(L)\vert_{u=\gbar^2(L),m(L)=0},
\end{equation} the expansion is known to second order,
\begin{equation} 
\label{eq:omega_pt}
\omega(u) = v_1 + v_2\,u + \rmO(u^2),
\end{equation} where the coefficients for $\Nf=3$ evaluate to
  \begin{equation}
     4\pi v_1   = 1.797887(5)\,, \qquad
    (4\pi)^2v_2 = -0.741(14)\,.
  \label{eq:v_12}
\end{equation}  Starting from
\begin{equation} \bar{g}_\nu^2(L) = \bar{g}^2(L)
\left[1-\nu\bar{g}^2(L) \omega\left(\gbar^2(L)\right)\right]^{-1}\,,
  \label{eq:gnusq}
\end{equation} we obtain the 2-loop relation,
\begin{equation} 
\alpha_\nu(\mu) = \alpha(\mu) + (4\pi v_1\nu) \alpha^2(\mu)
+ (4\pi)^2 \left(v_2\nu+ v_1^2\nu^2\right) \alpha^3(\mu) + \ldots\,.
\end{equation} Inverting perturbatively and combining with the
previous equations we have
\begin{equation} 
\alpha_\MSbar^{}(s\mu) = \alpha_\nu^{}(\mu) +
c_{1}^{\nu}(s) \alpha_\nu^2(\mu) + c_{2}^{\nu}(s) \alpha_\nu^3(\mu) +
\ldots\,,
\label{eq:alphaMSbar-alphaSFnu}
\end{equation} 
where
\begin{eqnarray} c_1^{\nu}(s) &=& c_1(s) -4\pi v_1 \nu\,, \\
c_2^{\nu}(s) -  \left(c_1^{\nu}(s)\right)^2 &=& c_2(s) - \left(c_1(s)\right)^2  - (4\pi)^2 v_2 \nu\,.
\end{eqnarray} 
In the perturbative matching of couplings one
occasionally applies the principle of ``fastest apparent
convergence'', which implies that $s=s^\star$ is chosen such as to make
the one-loop coefficient, $c_1^\nu(s^\star)$, vanish.  This is the case
for
\begin{equation} 
  \ln(s^\star) = \dfrac{c_1^\nu(1)}{8\pi b_0} = \frac{2\pi}{9} c_1^\nu(1)\,,
  \label{eq:scale_ratio}
\end{equation} and with this choice one obtains the relation,
\begin{equation} 
\alpha_\MSbar(s^\star \mu) = \alpha^{}_\nu(\mu) +
c_2^{\nu}(s^\star) \alpha_\nu^3(\mu) + \rmO\left(\alpha_\nu^4\right)\,.
\label{eq:FAP}
\end{equation}

\subsection{Perturbation theory and the $\Lambda$-parameter}

There are various ways to define a target precision for
$\alpha_s$. Instead of referring to the coupling in some scheme at
some scale it is attractive to instead refer to the
$\Lambda$-parameter. Given the coupling $\gbar_{\rm x}(L)$ in a scheme
${\rm x}$ the $\Lambda$-parameter in this scheme is a special solution
of the Callan--Symanzik equation of the form
\begin{eqnarray} \Lambda^{}_{\rm x} &=& L^{-1} \varphi^{}_{\rm
x}(\gbar^{}_{\rm x}(L))\,,
\end{eqnarray} with
\begin{eqnarray} \varphi_{\rm x}(\gbar) = ( b_0 \gbar^2
)^{-b_1/(2b_0^2)} \rme^{-1/(2b_0 \gbar^2)} \label{e:phig} \times
\exp\left\{-\int\limits_0^{\gbar} \rmd g\ \left[\frac{1}{\beta_{\rm
x}(g)} +\frac{1}{b_0g^3} - \frac{b_1}{b_0^2g} \right] \right\}.
\end{eqnarray} 
Note that this definition is independent of
perturbation theory provided the coupling and its $\beta$-function are
defined non-perturbatively.  In practice, however, one would like to
evaluate the $\Lambda$-parameter at a large energy scale $\mu=1/L$
such that the integral in the exponent can be safely evaluated in
perturbation theory.  The exact scheme-dependence of the
$\Lambda$-parameter is obtained by the one-loop matching of the
respective couplings. Labelling the schemes by ${\rm x}$ and ${\rm
y}$,
\begin{equation} {\bar g}^2_{\rm x}(L) = {\bar g}^2_{\rm y}(L) + c_{\rm xy}^{}{\bar g}_{\rm y}^4(L)
+ \ldots,
\end{equation} one obtains the {\em exact} relationship
\begin{equation} \Lambda_{\rm x}/\Lambda_{\rm y} = \rme^{c_{\rm
xy}^{}/2b_0}\,.
\end{equation} Note that this allows one to indirectly define
$\Lambda_\MSbar$ non-perturbatively, thereby justifying its use as a
reference definition.  With the perturbative matching coefficients of
the previous subsection we obtain the relationships (for $\Nf=3$),
\begin{equation} \Lambda/\Lambda_\MSbar^{} =0.38286(2)\qquad
\Lambda_\nu^{}/\Lambda = \exp\left(\nu\times 1.255162(4)\right),
  \label{eq:Lambda_Lambdanu}
\end{equation} where $\Lambda$ and $\Lambda_\nu$ are the parameters
for the SF and SF$_\nu$ scheme, respectively.  In particular, the
ratio $s^\star$ of scales used in Eq.~(\ref{eq:scale_ratio}) is given
by the ratio of the respective $\Lambda$-parameters.

\subsection{On exponentially suppressed corrections to perturbation
theory}

The perturbative expansion of the path integral generates an asymptotic series, 
with zero radius of convergence. In applications one then hopes 
that, for the accessible range of couplings, the perturbative series
provides a good quantitative description of the observable. 
The observables we consider here, the couplings in the SF$_\nu$ schemes, 
are defined non-perturbatively in Euclidean space-time, with an infrared cutoff provided by
the finite space-time volume. These properties are advantageous for
perturbation theory, in particular, there should be no renormalon
problem~\cite{Beneke:1998ui}.  
Lattice QCD provides very good non-perturbative control of these observables,
for couplings $\alpha$ in the range $0.1-0.2$ (cf.~Section~3).
Before testing perturbation theory, we would like to identify  
exponentially suppressed terms in the coupling which might preclude 
a good quantitative description of the non-perturbative data.
Such terms are associated with local minima of the action, e.g.~those corresponding to
the classical solutions of the field equations. Given the instanton
bound, Eq.~(\ref{eq:instbd}), and the absolute minimum $S_g[B]=2\pi^2/g_0^2$ of the
action~(Eq.~(\ref{eq:Sclass}), with $\eta=0$), 
contributions from the $|Q|=1$ instanton sector to our observables are 
accompanied by a suppression factor $\exp(-6\pi^2/g_0^2)=\exp(-3\pi/(2\alpha))$ 
and are therefore numerically irrelevant for our range of couplings. We may then ask the question
whether there are further secondary minima of the action
which are less strongly suppressed. Hence we are looking
for a secondary minimum $B^\ast_\mu$ of the gauge action in the $Q=0$ sector,
which satisfies
\begin{equation}
   \Delta S = S_g[B^\ast] - S_g[B] < 6\pi^2/g_0^2\,.
\end{equation} 
In fact there are ``large'' gauge transformations at $x_0=0$ corresponding to gauge
functions $\omega(\bfx)$ which are topologically trivial but are not
subject to the gauge fixing procedure around $B_\mu$~\cite{Luscher:1992an}.  
In order to find potential secondary minima we have resorted to a numerical experiment in the
lattice discretized theory. More precisely, we have first performed numerical simulations of the 
pure SU(3) Yang-Mills theory on a lattice with linear extent $L/a=8$, (at $\beta\equiv 6/g_0^2= 5.7$), 
and generated a long Monte Carlo history of  about $64.000$ configurations, 
corresponding to 128.000 MDU, using
the same simulation code as for our $\Nf=3$ QCD simulations (cf.~Section~3). Every 5$^\text{th}$ 
gauge configuration has been taken as initial condition for the gradient flow equation~\cite{Luscher:2010iy},
which we then integrated up to very large flow times $t$, corresponding
to $c=\sqrt{8t}/L =10$; the gradient flow is a smoothing operation and drives the gauge field
towards a local minimum of the action. At large flow times we selected the gauge field configurations 
in the $Q=0$ sector\footnote{In practice we defined this to mean gauge configurations 
for which $|Q|<0.5$, with $Q$ defined as in ref.~\cite{Luscher:2010iy}.}. 
Apart from the background field, Eq.~(\ref{eq:contBF}), 
we have indeed found a single further local minimum. In order to check
for its stability and  
to obtain its continuum limit, we have performed similar simulations
on finer lattices  
with $L/a=12,16,24$, and bare couplings such as to keep $\gbar^2(L/2)
=2.77$ approximately fixed.  
After extrapolation to the continuum and in the temporal gauge we find that 
this secondary minimum corresponds to the spatially constant Abelian field,
\begin{eqnarray} B^\ast_1(x) &=& \dfrac{i\pi}{L}\left\{{\rm
diag}\left(-\frac13,\frac13,0\right) +\frac{x_0}{T}{\rm
diag}\left(-\frac23,0,\frac23\right)\right\}\,,\\ B^\ast_2(x) &=&
B_1(x)\,,\\ B^\ast_3(x) &=& \dfrac{i\pi}{L}\left\{{\rm
diag}\left(-\frac73,\frac13,2\right) +\frac{x_0}{T}{\rm
diag}\left(-\frac43,0,\frac43\right)\right\}\,.
\end{eqnarray} The boundary conditions at $x_0=0$ thus are given as
\begin{equation} B^\ast_1(0,\bfx) = B^\ast_2(0,\bfx) = C_1^\omega =
C_2^\omega= \dfrac{i\pi}{L}{\rm diag}\left(-\frac13,\frac13,0\right),
\end{equation} and
\begin{equation} B^\ast_3(0,\bfx) = C_3^\omega = \dfrac{i\pi}{L}{\rm
diag}\left(-\frac73,\frac13,2\right)\,.
\end{equation} The gauge function $\omega(\bfx)$ is thus non-constant
in the $x_3$-direction, which induces the shift by $\pm 2\pi$ in 2 of
the angles of $C_3^\omega$, in addition to the permutation of the
colour 2- and 3-components of $\phi$, Eq.~(\ref{eq:phi_phiprime}).
Obviously the spatial directions can be permuted, so this minimum has a 3-fold degeneracy.
Hence, the classical field $B^\ast_\mu$ is Abelian and spatially
constant, but with boundary values, transformed by the
gauge function
\begin{equation} \omega(\bfx) = \begin{pmatrix}
\exp\left(i\frac{2\pi}{L}x_3\right) & 0 & 0 \cr 0 & 0 & -1\cr 0 &
\exp\left(-i\frac{2\pi}{L}x_3\right) & 0\cr
                    \end{pmatrix}\,.
\label{eq:omegxa_Bstar}
\end{equation} 
To find the gap in the gauge action we insert the
non-zero components of the field tensor
\begin{equation} G^\ast_{0k} = \partial_0 B_k^\ast =
\dfrac{i\pi}{L^2}\times
\begin{cases} {\rm diag}\left(-\frac23,0,\frac23\right), &\text{if
$k=1,2$},\\ {\rm diag}\left(-\frac43,0,\frac43\right), & \text{if
$k=3$},
\end{cases}
\end{equation} into the gauge action~Eq.~(\ref{eq:S_g}), with the
result
\begin{equation} g_0^2 S[B^\ast] = -L^4
\sum_{k=1}^3\tr\{G^\ast_{0k}G^\ast_{0k}\} = \dfrac{16\pi^2}{3}\,.
\end{equation} Hence the gap, $\Delta S$, is found to be $10/3$ in
units of $\pi^2/g_0^2$ which is $2/3$ below the $Q=1$ instanton
threshold. This leads to a suppression factor $\exp(-g_0^2 \Delta
S/(4\pi\alpha)) = \exp(-5\pi/(6\alpha))$.  For the range of couplings in our study,
this factor varies from a few times $10^{-6}$ to below $10^{-10}$, which
renders such a non-perturbative contribution completely negligible.

\section{Lattice set-up and simulations}
\label{sec:lattice}
In this section we briefly describe the main elements of the
lattice set-up chosen for this study and discuss some details
pertaining to the error treatment.

\subsection{Lattice action} \label{subsec:ctctt}

We choose the standard Wilson plaquette action for the gauge fields and
three flavours of non-perturbatively O($a$) improved Wilson fermions. The lattice
action is then given by $S=S_g+S_f$, with
\begin{eqnarray}
  \label{e_Sg}
  S_g[U] &=& \frac{1}{g_0^2}\sum_{p}w(p)\,\tr\{1-U(p)\} ,\label{eq:Slat_g}\\[1ex]
  S_f[U,\bar\psi,\psi] &=& a^4\sum_{x}\bar\psi(x)(D+\delta D_b + m_0)\psi(x) . \label{e_Sf}
\end{eqnarray}  
The gauge field action $S_g$ is a sum over all oriented 
plaquettes $p$ on the lattice, with the weights $w(p)$, and 
the parallel transporters $U(p)$  around $p$. With the gauge field
boundary conditions given in terms of the Abelian fields, Eq.~(\ref{eq:bc_phases}),
\begin{equation}
 U_k(0,\bfx) = \exp(a C^{}_k),\qquad U_k(L,\bfx) = \exp(a C'_k),
 \label{eq:bcgauge}
\end{equation}
the gauge part of the action is completely specified by setting $w(p)=1$
except for timelike plaquettes touching one of the boundaries for which $w(p)=\ct$.
The Dirichlet boundary conditions for the quark fields look exactly
the same as in the continuum, cf.~Section~2. Like in the continuum 
we take the fermionic fields to be spatially periodic and implement the phase $\theta=\pi/5$ via a constant
U(1) background field $\lambda_\mu^{} = \exp(ia\theta_\mu/L)$, with $\theta_\mu = (1-\delta_{\mu,0})\theta$. 
With the covariant derivatives,
\begin{eqnarray}
  \nabla_\mu^{}\psi(x)     &=& \frac1a [\lambda_\mu U(x,\mu)\psi(x+a\hat\mu)-\psi(x)],\\
  \nabla_\mu^{\ast}\psi(x) &=&
   \frac1a [\psi(x)-\lambda_\mu^\ast U(x-a\hat\mu,\mu)^{\dag}\psi(x-a\hat\mu)]\,,
\end{eqnarray}
the Wilson-Dirac operator in the fermionic action (\ref{e_Sf}) takes the form,
\begin{equation}
  D = \frac12 \sum_{\mu=0}^3
  \left\{\gamma_\mu(\nabla_\mu^\ast+\nabla_\mu^{})- a\nabla_\mu^\ast\nabla_\mu^{}\right\}
 + \csw\,\frac{ia}{4}\sum_{\mu,\nu=0}^3
 \,\sigma_{\mu\nu}{\cal F}_{\mu\nu} \label{eq:Dlat}\,,
\end{equation}
which includes the Sheikholeslami-Wohlert term~\cite{Sheikholeslami:1985ij}. For
the clover leaf definition of the field strength tensor, ${\cal F}_{\mu\nu}$, 
we refer to~\cite{Luscher:1996sc} and the improvement coefficient $\csw(g_0)$ 
is set non-perturbatively using the result from \cite{Yamada:2004ja}. 
Finally, the fermionic $\rmO(a)$ boundary counterterm action is specified by \cite{Luscher:1996sc}
\begin{equation}
  \label{eq:Db}
  \delta D_b \psi(x) = (\ctt-1)(\delta_{x_0,a}+\delta_{x_0,T-a}) \psi(x)\,.
\end{equation}
The 2 boundary counterterm coefficients, $\ct(g_0)$ and $\ctt(g_0)$ are 
set to their perturbative two- and one-loop expressions, respectively~\cite{Luscher:1993gh,Bode:1999sm},
\begin{eqnarray}
 \label{eq:ct}
 \ct(g_0) &= &  1 + \ct^{(1)}g_0^2 + \ct^{(2)}g_0^4 + \rmO(g_0^6)\,,\\
 \label{eq:cttilde}
\ctt(g_0) &= &  1 + \ctt^{(1)}g_0^2 + \rmO(g_0^4)\,,
\end{eqnarray}
with the known perturbative coefficients for $N=3$ colours given by
\begin{eqnarray}
  \ct^{(1)} &=&  -0.0890 + 0.019141\times \Nf \hspace*{2.25cm} 
  \quad{\buildrel{\scalebox{0.7}{$\Nf=3$}}\over{=}}\quad -0.0315\,,\\[1ex]
  \ct^{(2)} &=&  -0.0294 + 0.002\times \Nf + 0.000(1)\times \Nf^2 
  \quad{\buildrel{\scalebox{0.7}{$\Nf=3$}}\over{=}}\quad -0.0234\,,\\
 \ctt^{(1)} &=&  -0.01795\,.
\end{eqnarray}
We notice a significant cancellation in the one-loop
term $\ct^{(1)}$ between the gluon and fermion contributions. We interpret the resulting
relative size of one- and two-loop terms for $\Nf=3$ as an accident 
and not a sign for a poor behaviour of the series in general.

\subsection{Lattice observables}

Like in the continuum,  the basic observables $1/\gbar^2$ and $\vbar$ are 
given as expectation values, Eq.~(\ref{eq:expectval}), of gauge invariant fields, 
which are now obtained as $\eta$- and $\nu$-derivatives of the lattice 
action\footnote{The fermionic action depends on $\eta$ through the Sheikholeslami-Wohlert 
term in Eq.~(\ref{eq:Dlat}) and thus also contributes to the observable, cf.~appendix~A 
of ref.~\cite{DellaMorte:2004bc} for details.}~(\ref{eq:Slat_g},\ref{e_Sf}). 
The lattice version of the Abelian background field
takes the form,
\begin{equation}
  V_\mu(x) = \exp\left(a B_\mu(x)\right),
\end{equation}
with $B_\mu(x)$ the continuum expression, Eq.~(\ref{eq:contBF}).
Cutoff effects with such Abelian gauge fields are known to be small~\cite{Luscher:1992an}. 
Indeed, the $\eta$-derivative of $S_g[V]$ yields the lattice renormalization constant 
\begin{equation}
  k =  12 (L/a)^2 [ \sin(\gamma)+\sin(2\gamma)], \qquad \gamma=\frac13 \pi (a/L)^2\,,
\end{equation}
which converges to $12\pi$ with O($a^4$) corrections. 
We will use this lattice definition of $k$ in order to ensure $\bar{g}^2=g_0^2$ {\em exactly} at lowest order.
Note that this also holds for $\gbar_\nu^2$, since $\vbar$ vanishes identically at tree level.

On the lattice with Wilson quarks, the chiral limit is not sharply defined, and one
also needs to specify the exact definition used. For given bare coupling $g_0$, we require
the PCAC quark mass,
\begin{equation}
   m(L) = \left.\frac{\frac12(\partial_0^{}+\partial_0^\ast) f_{\rm A}(x_0) + 
   c_{\rm A}(g_0) a \partial_0^\ast\partial_0^{} f_{\rm P}(x_0)}  {2 f_{\rm P}(x_0)}\right\vert_{x_0=L/2}\,,
   \label{eq:mPCAC}
\end{equation}
to vanish on an $(L/a)^4$ lattice with the Abelian boundary conditions, Eq.~(\ref{eq:bcgauge}).
Here $f_{\rm A}(x_0)$ and $f_{\rm P}(x_0)$ are Schr\"odinger functional correlation functions defined 
e.g.~in Eqs.~(2.1) and~(2.2) of \cite{Luscher:1996ug},
and $\partial_0$, $\partial_0^\ast$ are the forward and backward lattice time derivatives, 
respectively. Finally, the improvement coefficient, $c_{\rm A}$, is set to its 
perturbative 1-loop value~\cite{Luscher:1996vw,Sint:1997jx}, since a non-perturbative estimate 
is not available for $\Nf=3$ and our choice of gauge action. Given that
we do not attempt to reach the low energy, hadronic regime, we expect one-loop perturbation theory to work reasonably
well for $c_{\rm A}$. The chiral limit is now defined by $m(L)=0$, and, for given bare coupling 
$g_0^2\equiv 6/\beta$, the bare mass $m_0$ for which this equation holds, 
defines the critical mass parameter or, equivalently, 
the critical $\kappa$,
\begin{equation}
  am_0 = am_{\rm cr}(g_0) = 1/\left(2\kappa_{\rm cr}(g_0)\right)-4\,.
\end{equation}
With these conventions we may now define the lattice observables. Specifying the value $u$ of the coupling $\gbar^2(L)$ 
at vanishing quark mass $m(L)$ defines our approach to the continuum limit, and other lattice observables 
are then well-defined functions of $u$. In particular $\vbar$ gives rise to 2 lattice observables
\begin{eqnarray}
  \Omega(u,a/L) &=& \left.\vbar(L)\right\vert_{u=\bar{g}^2(L),m(L)=0}\,,\\ 
  \tilde{\Omega}(u,a/L) &=& \left.\vbar(L)\right\vert_{u=\bar{g}^2(L),m(L/2)=0} \,,
\end{eqnarray}
which differ by the chiral limit definition. The appearance of 2 lattice versions for $\omega(u)$ 
is a consequence of the definition  of the lattice step-scaling functions  through
\begin{equation}
  \Sigma(u,a/L) = \left.\bar{g}^2(2L)\right\vert_{\bar{g}^2(L)=u,m(L)=0}\,,
 \label{eq:SSFlat}
\end{equation}
which requires simulations on lattices with resolutions $L/a$ and $2L/a$, at
the same bare parameters. In particular, the simulations on the $2L/a$-lattices 
are performed at the bare mass parameters for which the PCAC mass vanishes on the $L/a$ lattice. 
Finally, we also consider the lattice step-scaling functions for $\gbar^2_\nu$,  
\begin{equation}
    \Sigma_\nu(u,a/L) = \left.\bar{g}_\nu^2(2L)\right\vert_{\bar{g}_\nu^2(L)=u,m(L)=0}\,,
\end{equation}
at non-zero values of $\nu$.

\subsection{Perturbatively improved lattice observables} \label{subsec:pertimp}

In order to accelerate the approach to the continuum limit one may
use perturbation theory to subtract the lattice artefacts order by order
in the coupling from the non-perturbative data~\cite{deDivitiis:1994yz}. The 2-loop calculation
in \cite{Bode:1999sm} has been carried out in the very same lattice regularized theory,
and the two-loop lattice artefacts in the $\nu=0$ step-scaling functions,
\begin{equation}
       \delta(u,a/L) = \frac{\Sigma(u,a/L) -\sigma(u)}{\sigma(u)}
                     = \delta_1(a/L) u + \delta_2(a/L) u^2 + \rmO(u^3)\,,
   \end{equation}
are indeed available to this order. With the coefficients for $\Nf=3$ from table~\ref{tab:pertimp},   
one may thus define the improved step-scaling functions,
\begin{equation}
   \Sigma^{(i)}(u,L/a) = \frac{\Sigma(u,L/a)}{1+ \sum_{k=1}^i \delta_k(L/a) u^k},
\end{equation}
up to loop order $i=2$. By construction, the leading cutoff effects for $i=0,1,2$ are then given by\footnote{Here
and in the following the $i=0$ label refers to unimproved data, for instance $\Sigma^{(0)}(u,a/L) = \Sigma(u,a/L)$, etc.}
\begin{equation}
   \Sigma^{(i)}(u,a/L) = \sigma(u) + \frac{a}{L}\times \rmO(u^4)  +\frac{a^2}{L^2}\times\rmO(u^{2+i})\,,
\end{equation}
and are thus suppressed by additional powers of the coupling. The term linear in $a/L$ is due to the incomplete
cancellation of the O($a$) boundary effects and could be eliminated by a non-perturbative determination
of $\ct$ and $\ctt$. We will come back to the question 
of remnant O($a$) effects in Subsect.~\ref{subsec:oamodel}.

For the observables $\Omega$ and $\tilde{\Omega}$ one parametrizes the cutoff effects by
2 functions, $\epsilon$ and $\tilde{\epsilon}$. For $\Omega$ we have
\begin{eqnarray}
  \Omega(u,a/L) &=& \omega(u)\left[1+\epsilon(u,a/L)\right]\,,
\end{eqnarray}
with perturbative expansion
\begin{equation}
   \epsilon(u,a/L) = \epsilon_1(a/L) + \epsilon_2(a/L) u + \rmO(u^2),
\end{equation}
and analogous equations hold for $\tilde \Omega$ and $\tilde{\epsilon}$. 
Unfortunately, the published results of the 2-loop calculation
do not allow for the extraction of the cutoff effects for this case, so that 
the perturbatively improved observables,
\begin{eqnarray}
         \Omega^{(i)}(u,a/L) &=& \frac{\Omega(u,a/L)}{1+\sum_{k=1}^i\epsilon_k(a/L) u^{k-1}}\,,\\
 \tilde{\Omega}^{(i)}(u,a/L) &=& \frac{\tilde{\Omega}(u,a/L)}{1+\sum_{k=1}^i\tilde{\epsilon}_k(a/L) u^{k-1}}\,,
\end{eqnarray}
are only available to 1-loop order, $i=1$, with the coefficients $\epsilon_1$ and $\tilde{\epsilon}_1$ 
given in table~\ref{tab:pertimp}. 

The same remark applies to the step-scaling function $\Sigma_\nu$ for $\nu\ne 0$.
Using the notation,
\begin{equation}
   \delta_\nu(u,a/L) = \delta_1^\nu(a/L) u + \rmO(u^2),  
\end{equation}
the one-loop coefficient is given by
\begin{equation}
  \delta_1^\nu(L/a) = \delta_1(L/a) + \nu v_1 \left[\tilde{\epsilon}_1(a/2L) - \epsilon_1(a/L)\right],
  \label{eq:del1nu}
\end{equation}
where $v_1$ is the expansion coefficient of the continuum function $\omega(u)$, Eq.~(\ref{eq:v_12}).
Values for $\delta_1^\nu$ can be inferred from table~\ref{tab:pertimp}, for $\Nf=3$ and 
the lattice sizes relevant for this study.
\begin{table}[h]
\centering
  \begin{tabular}{CRRCRC}\toprule
      L/a & \delta_1 \times 10^2  & \delta_{2}\times 10^2 & 
      (\delta^\nu_{1}-\delta_1)/\nu \times 10^2 
      & \epsilon_1\times 10^2 & \tilde{\epsilon}_1(a/2L)\times 10^2 \\\midrule
  4 &  -1.02700 &  0.28560 & -3.94211 & 33.26842 & 5.71494 \\
  6 &  -0.43600 &  0.02510 & -1.44433 & 12.20048 & 2.10529 \\
  8 &  -0.22700 & -0.01380 & -0.61453 & 5.42725  & 1.13201 \\
 10 &  -0.13800 & -0.01260 & -0.32597 & 2.99507  & 0.71670 \\
 12 &  -0.09400 & -0.00960 & -0.20334 & 1.91825  & 0.49698 \\
    \bottomrule
  \end{tabular}
   \caption{Values of the coefficients for $\Nf=3$ and the relevant lattice sizes, as required 
    for perturbative cancellation of lattice artefacts up to 2-loop order in $\Sigma$, and to one-loop order
    in $\Sigma_\nu$, $\Omega$ and $\tilde{\Omega}$, cf.~text.}
\label{tab:pertimp}
\end{table}

\subsection{Simulation parameters and statistics}

Using the openQCD code~\cite{Luscher:2012av,openqcd:2013} we have simulated lattice sizes $L/a=4,6,8,10,12$
around 9 values of the coupling $\bar{g}^2(L)=u$ in the range $1.1-2.0$, cf.~table~\ref{tab:rawdata1}.
At the same bare coupling $g_0^2=6/\beta$ and bare quark mass $am_0 = 1/(2\kappa)-4$ we then doubled the lattice
sizes and simulated for $2L/a=8,12,16$ and, in 3 cases also for 
$2L/a=24$, cf.~table~\ref{tab:rawdata2}. 
Starting from the $L/a=12$ lattices we have tried to approximately match the values
of the coupling for $\nu=0$ at $L/a=4,6,8$, so as to be able to do continuum extrapolations of the step-scaling function 
at individual values of the coupling, without the necessity for large interpolations of
the data. 

As a target precision we chose the criterion,
\begin{equation}
  \Delta\left(\frac{1}{\gbar^2}\right) = \frac{\Delta \gbar^2}{\gbar^4} \approx 0.001\,,
  \label{eq:target-prec}
\end{equation}
which is reached for most of our data except for some $L/a=10$ lattices. These lattices
were however not used for the step scaling procedure as we did not generate corresponding configurations
on $2L/a=20$ lattices. Except for some checks we also refrained from 
using lattices as small as $L/a=4$ and thus do not list the results here. 
However, the $L/a=10$ data and the $2L/a=8$ data are
used for the continuum extrapolation of $\Omega$ and $\tilde{\Omega}$, respectively,
and are therefore included in the tables.

\begin{table}[t]
  \small
  \centering
  \begin{tabular}[c]{ccccccccc}\toprule
   $L/a$& $\beta$& $\kappa$   &$\gbar^2$& $\Delta\gbar^2$ 
                                                 & $\gbar^2_{\nu=0.3}$ 
                                                          & $\Delta\gbar_{\nu=0.3}^2$ 
                                                                   &$\vbar$ & $\Delta\vbar$ \\\midrule
      6 & 6.2650 & 0.13558688  & 2.0194 & 0.0032 & 2.1991 & 0.0042 & 0.1349 & 0.0016        \\ 
      6 & 6.5964 & 0.13499767  & 1.7983 & 0.0025 & 1.9448 & 0.0033 & 0.1396 & 0.0017        \\ 
      6 & 6.9283 & 0.13444591  & 1.6247 & 0.0021 & 1.7462 & 0.0029 & 0.1427 & 0.0018        \\ 
      6 & 7.2604 & 0.13393574  & 1.4799 & 0.0015 & 1.5831 & 0.0020 & 0.1467 & 0.0016        \\ 
      6 & 7.5769 & 0.13348828  & 1.3680 & 0.0011 & 1.4568 & 0.0015 & 0.1485 & 0.0014        \\ 
      6 & 7.8935 & 0.13307660  & 1.2703 & 0.0009 & 1.3487 & 0.0013 & 0.1526 & 0.0014        \\ 
      6 & 8.2103 & 0.13269801  & 1.1864 & 0.0009 & 1.2552 & 0.0012 & 0.1540 & 0.0015        \\ 
      6 & 8.5271 & 0.13234995  & 1.1125 & 0.0008 & 1.1723 & 0.0011 & 0.1528 & 0.0016        \\ 
      8 & 6.4575 & 0.13525498  & 2.0201 & 0.0034 & 2.1875 & 0.0046 & 0.1262 & 0.0017        \\ 
      8 & 6.7900 & 0.13467912  & 1.7943 & 0.0031 & 1.9341 & 0.0043 & 0.1343 & 0.0022        \\ 
      8 & 7.1225 & 0.13414622  & 1.6173 & 0.0025 & 1.7291 & 0.0034 & 0.1332 & 0.0021        \\ 
      8 & 7.4550 & 0.13365676  & 1.4783 & 0.0019 & 1.5728 & 0.0026 & 0.1354 & 0.0020        \\ 
      8 & 7.7721 & 0.13322862  & 1.3629 & 0.0016 & 1.4440 & 0.0021 & 0.1374 & 0.0020        \\ 
      8 & 8.0891 & 0.13283568  & 1.2657 & 0.0009 & 1.3374 & 0.0013 & 0.1412 & 0.0014        \\ 
      8 & 8.4062 & 0.13247474  & 1.1845 & 0.0008 & 1.2473 & 0.0011 & 0.1417 & 0.0014        \\ 
      8 & 8.7232 & 0.13214306  & 1.1122 & 0.0013 & 1.1674 & 0.0017 & 0.1419 & 0.0024        \\ 
     10 & 6.6046 & 0.13498493  & 2.0259 & 0.0124 & 2.1879 & 0.0167 & 0.1218 & 0.0059        \\ 
     10 & 6.6073 & 0.13498022  & 2.0129 & 0.0035 & 2.1803 & 0.0049 & 0.1271 & 0.0017        \\ 
     10 & 7.6010 & 0.13344250  & 1.4898 & 0.0069 & 1.5799 & 0.0090 & 0.1276 & 0.0067        \\ 
     10 & 7.6063 & 0.13343533  & 1.4794 & 0.0019 & 1.5719 & 0.0025 & 0.1326 & 0.0018        \\ 
     10 & 8.8675 & 0.13198989  & 1.1118 & 0.0030 & 1.1638 & 0.0040 & 0.1341 & 0.0056        \\ 
     10 & 8.8755 & 0.13198218  & 1.1093 & 0.0010 & 1.1633 & 0.0013 & 0.1395 & 0.0019        \\ 
     12 & 6.7300 & 0.13475901  & 2.0123 & 0.0037 & 2.1725 & 0.0048 & 0.1221 & 0.0018        \\ 
     12 & 7.7300 & 0.13326291  & 1.4805 & 0.0020 & 1.5752 & 0.0026 & 0.1355 & 0.0019        \\ 
     12 & 9.0000 & 0.13185703  & 1.1089 & 0.0014 & 1.1614 & 0.0018 & 0.1358 & 0.0024        \\ 
    \bottomrule
  \end{tabular}
  \caption{Simulation parameters and results on the $L$-lattices. 
 The hopping parameter $\kappa$ was tuned such that the PCAC mass $m(L)$, Eq.~(\ref{eq:mPCAC}),
 vanishes.}
  \label{tab:rawdata1}
\end{table}

\begin{table}[t]
  \small
  \centering
  \begin{tabular}[c]{ccccccccc}\toprule
  $L/a$&$\beta$ & $\kappa$    &$\gbar^2$& $\Delta\gbar^2$ 
                                                & $\gbar^2_{\nu=0.3}$ 
                                                         & $\Delta\gbar_{\nu=0.3}^2$ 
                                                                  & $\vbar$& $\Delta\vbar$ \\\midrule
     8 & 6.0522 & 0.13546638  & 2.4124 & 0.0044 & 2.6281 & 0.0057 & 0.1134 & 0.0014        \\ 
     8 & 6.3757 & 0.13492039  & 2.0955 & 0.0039 & 2.2648 & 0.0052 & 0.1189 & 0.0018        \\ 
     8 & 6.7145 & 0.13437600  & 1.8586 & 0.0035 & 1.9989 & 0.0047 & 0.1259 & 0.0021        \\ 
     8 & 7.0275 & 0.13390509  & 1.6756 & 0.0020 & 1.7919 & 0.0026 & 0.1291 & 0.0015        \\ 
     8 & 7.3496 & 0.13345482  & 1.5286 & 0.0018 & 1.6277 & 0.0024 & 0.1328 & 0.0017        \\ 
     8 & 7.6782 & 0.13303090  & 1.3997 & 0.0013 & 1.4855 & 0.0018 & 0.1376 & 0.0015        \\ 
     8 & 7.9822 & 0.13266902  & 1.3014 & 0.0011 & 1.3743 & 0.0015 & 0.1359 & 0.0015        \\ 
     8 & 8.3130 & 0.13230601  & 1.2128 & 0.0010 & 1.2784 & 0.0013 & 0.1410 & 0.0016        \\ 
    12 & 6.2650 & 0.13558688  & 2.4568 & 0.0060 & 2.6788 & 0.0081 & 0.1124 & 0.0018        \\ 
    12 & 6.5964 & 0.13499767  & 2.1287 & 0.0042 & 2.2995 & 0.0054 & 0.1163 & 0.0018        \\ 
    12 & 6.9283 & 0.13444591  & 1.8780 & 0.0029 & 2.0175 & 0.0039 & 0.1227 & 0.0017        \\ 
    12 & 7.2604 & 0.13393574  & 1.6839 & 0.0024 & 1.8045 & 0.0033 & 0.1323 & 0.0019        \\ 
    12 & 7.5769 & 0.13348828  & 1.5378 & 0.0019 & 1.6370 & 0.0026 & 0.1314 & 0.0018        \\ 
    12 & 7.8935 & 0.13307660  & 1.4148 & 0.0016 & 1.5010 & 0.0021 & 0.1353 & 0.0018        \\ 
    12 & 8.2103 & 0.13269801  & 1.3114 & 0.0017 & 1.3860 & 0.0022 & 0.1369 & 0.0022        \\ 
    12 & 8.5271 & 0.13234995  & 1.2210 & 0.0014 & 1.2880 & 0.0019 & 0.1420 & 0.0022        \\ 
    16 & 6.4575 & 0.13525498  & 2.4540 & 0.0056 & 2.6708 & 0.0072 & 0.1103 & 0.0016        \\ 
    16 & 6.7900 & 0.13467912  & 2.1251 & 0.0043 & 2.2970 & 0.0057 & 0.1174 & 0.0018        \\ 
    16 & 7.1225 & 0.13414622  & 1.8810 & 0.0039 & 2.0230 & 0.0051 & 0.1244 & 0.0021        \\ 
    16 & 7.4550 & 0.13365676  & 1.6863 & 0.0029 & 1.8017 & 0.0039 & 0.1265 & 0.0021        \\ 
    16 & 7.7721 & 0.13322862  & 1.5375 & 0.0022 & 1.6370 & 0.0029 & 0.1317 & 0.0019        \\ 
    16 & 8.0891 & 0.13283568  & 1.4164 & 0.0018 & 1.5011 & 0.0024 & 0.1328 & 0.0020        \\ 
    16 & 8.4062 & 0.13247474  & 1.3090 & 0.0017 & 1.3825 & 0.0022 & 0.1353 & 0.0021        \\ 
    16 & 8.7232 & 0.13214306  & 1.2204 & 0.0014 & 1.2842 & 0.0019 & 0.1358 & 0.0021        \\ 
    24 & 6.7300 & 0.13475901  & 2.4517 & 0.0067 & 2.6732 & 0.0087 & 0.1126 & 0.0019        \\ 
    24 & 7.7300 & 0.13326291  & 1.6847 & 0.0033 & 1.7980 & 0.0042 & 0.1246 & 0.0023        \\ 
    24 & 9.0000 & 0.13185703  & 1.2232 & 0.0022 & 1.2892 & 0.0029 & 0.1394 & 0.0032        \\ 
    \bottomrule
  \end{tabular}
  \caption{Simulation parameters and results on the doubled lattices. 
 The hopping parameter $\kappa$ was tuned such that the PCAC mass $m(L/2)$
 vanishes, cf.~Eq.~(\ref{eq:mPCAC}).}
  \label{tab:rawdata2}
\end{table}

Note that the choice of the reference value $\nu_0=0.3$ is rather arbitrary. In fact, the data
in the table for $\gbar^2$, $\gbar^2_{\nu_0=0.3}$ and $\vbar$, with their statistical
errors enables the reconstruction of the coupling at any
value of $\nu$, using Eq.~(\ref{eq:gnusq}) and straightforward error propagation,
\begin{equation}
  \frac{\Delta \gbar^2_\nu}{\gbar^4_\nu} = \left\{\frac{\nu}{\nu_0}
  \left(\frac{\Delta \gbar^2_{{\nu_0}}}{\gbar^4_{\nu_0}}\right)^2
  + \left(\frac{\Delta \gbar^2}{\gbar^4}\right)^2\left(1-\frac{\nu}{\nu_0}\right) 
  + \nu^2 \left(\Delta \vbar\right)^2\left(1-\frac{\nu_0}{\nu}\right)\right\}^{1/2}\,.
 \label{eq:var-nu}
  \end{equation}
We have checked that this reconstruction does indeed reproduce the result of a direct data analysis
at a given $\nu$-value, provided that the treatment of autocorrelations is done consistently 
for the couplings at all $\nu$-values and $\vbar$. We find that the precision for the $\nu=0$ coupling,
Eq.~(\ref{eq:target-prec}), translates to higher values for other choices of $\nu$, for instance
we find an increase of 20 percent for $\nu=0.3$ (from tables~\ref{tab:rawdata1} and \ref{tab:rawdata2}),
and ca.~50 percent for $\nu=-0.5$ from Eq.~(\ref{eq:var-nu}).

All statistical errors were determined using the 
$\Gamma$-method~\cite{Wolff:2003sm}. For our observables, one 
even has to be careful that one sums up the autocorrelation function
sufficiently far. Still the final autocorrelation
times range from values somewhat below 2 MDU for weak coupling and 
small $L/a$, to about 8 MDU at larger coupling and $L/a=24$.
Further details on the performance of our algorithms will be reported in~\cite{fritzsch:2018}.

\subsection{Treatment of statistical errors}

When forming the step scaling function $\Sigma(u,a/L)$ there are statistical uncertainties both for
 $\gbar^2(L)$, \tab{tab:rawdata1}, and for $\gbar^2(2L)$, \tab{tab:rawdata2}. 
These are propagated to the error of $\Sigma(u,a/L)$ with $u$ the central value of the estimate of $\gbar^2(L)$, via 
\begin{equation}
  (\Delta \Sigma(u,a/L))^2 = (\Delta \gbar^2(2L))^2 + \left(\frac{\partial \Sigma(u,a/L)}{\partial u} \Delta \gbar^2(L)\right)^2.
\label{eq:ssf-error}
  \end{equation}
To estimate the required derivative $\partial \Sigma/\partial u$ we differentiate the 
3-loop truncation of the continuum function, $\sigma(u)$, Eq.~(\ref{eq:sigcontpert}),
corrected for the known lattice artefacts at one- and two-loop order for $\nu\ne 0$ and $\nu=0$, respectively,
cf.~Subsect.~\ref{subsec:pertimp}. For $\nu=0$ this leads to
\begin{equation}
  \frac{\partial \Sigma}{\partial u} \approx 1 + 2(s_0+\delta_1) u + 3(s_1+\delta_2+s_0 \delta_1)u^2 + 4 s_2 u^3,
\end{equation}
and similarly for $\nu\ne 0$ with $\delta_1^\nu$ from Eq.~(\ref{eq:del1nu}), the unknown $\delta_2^\nu$ set to zero and with 
the scheme dependence of $s_2$ (via $b_2$, Eq.~(\ref{eq:s012})), taken into account. 
As a cross check, we also estimated the derivative directly from the data 
and found the differences to be negligible. 

For the study of the observables $\Omega$ and $\tilde\Omega$ we proceed similarly:
to obtain the derivative with respect to $u$ we first perform a
rough continuum extrapolation neglecting the errors on $u$. 
The resulting polynomial fit function
\begin{equation}
 \omega(u) \approx  0.14307  - 0.004693\times u + 0.0077906 \times u^2 -0.0105266 \times u^3 + 0.0023996 \times u^4\,,
 \label{eq:omegafit}
\end{equation}
is then differentiated to provide an estimate for $\partial\Omega/\partial u$ and $\partial\tilde \Omega/\partial u$,
neglecting any $L/a$-dependence of the derivative.

\subsection{Quality of tuning to the chiral limit}

An important aspect of Wilson fermions is the need to tune the bare quark mass parameter (parameterized by $\kappa$)
to a critical value, such that chiral symmetry is restored up to cutoff effects. For our choice of condition
$m(L)=0$, with the PCAC mass of Eq.~(\ref{eq:mPCAC}), we have performed extensive tuning runs
which enable a precision such that,
\begin{equation}
   |z| < 0.001\,, \qquad z= am(L)\times (L/a) = m(L)L,
 \label{eq:mass-prec}
\end{equation}
at all stages of the calculation~\cite{fritzsch:2018}. The corresponding values for $\kappa$ are
given in \tab{tab:rawdata1}. What is the tolerance of a slight mistuning of the mass?
Using 1-loop perturbative results from ref.~\cite{Sint:1995ch} for the mass dependence of
$\gbar^2$ and $\vbar$  we obtain, in the continuum limit,
\begin{equation}
  \left.\frac{\partial \gbar_\nu^2}{\partial z}\right\vert_{z=0} 
  = \Nf\times\left[0.0095683(1) -0.01418(5)\times\nu\right] \gbar^4_\nu\, + {\rm O}(\gbar^6_\nu)\,.
\end{equation}
This should be compared with the target statistical precision, which is, 
for $\nu=0$, given in Eq.~(\ref{eq:target-prec}).
We follow ref.~\cite{DellaMorte:2004bc} and allow for an uncertainty of about $1/3$
of the statistical error. Neglecting small cutoff effects in the mass derivative and for $\Nf=3$ 
this yields the bounds,
\begin{equation}
   |z| < \frac{(1/3)\times \Delta \gbar^2_\nu}{\left(\partial\gbar^2_\nu/\partial z\right)_{z=0}} \approx 
   \begin{cases} 0.010, & \text{$\nu=-0.5$\,,}\\
                 0.012, & \text{$\nu=0$\,,}\\
                 0.025, & \text{$\nu=0.3$\,,}
   \end{cases}              
\end{equation}
for the $\nu$-values that we chose for more detailed analysis in Sect.~\ref{sec:cont}.
We note that the achieved precision of the mass tuning, Eq.~(\ref{eq:mass-prec}),  
stays well within these bounds, by at least a factor 10. Even if these perturbative estimates
turned out to be significantly off the mark, e.g.~by a factor 2, 
the systematic error associated with imperfect quark mass tuning would still be negligibly small
and can thus be safely ignored.

\subsection{Lattice artefacts linear in $a/L$}
\label{subsec:oamodel}
Despite the use of a non-perturbatively O($a$) improved bulk action the
very presence of the time boundaries in the Schr\"odinger functional creates
lattice artefacts linear in $a$. In principle these could be cancelled by
an appropriate non-perturbative tuning of the improvement coefficients 
$\ct$ and $\ctt$, Eqs.~(\ref{e_Sg}, \ref{eq:Db}). In practice,
however, we are  
currently limited to the use of perturbative estimates,
Eqs.~(\ref{eq:ct}, \ref{eq:cttilde}).
Hence some remnant linear $a$-effects in our data cannot be excluded. 
Instead of including a corresponding term in the fit ansatz 
for the continuum extrapolations we try to estimate the size of
these uncertainties and include them as an additional systematic error.
Using a combination of simulations and perturbation theory we
have produced a model for the sensitivity of our data to 
a variation of $\ct$ and $\ctt$. The details are deferred to appendix~\ref{app:oamodel}, 
where we obtain linearized shifts of the data, for instance,
\begin{equation}
  \label{eq:Sigmact}
  \Sigma(u,a/L)\vert_{\ct' = \ct+\Delta\ct} = \Sigma(u,a/L)\vert_{\ct} + \Delta\ct \times \delta_{\ct}\Sigma(u,a/L),
\end{equation}
and analogously for a shift $\ctt'=\ctt+\Delta\ctt$. Hence, the model yields
an estimate of the data that would have been obtained if the 
simulations had been performed at slightly different values $\ct'$ and $\ctt'$. 
To complete the model we thus need an educated guess for
$\Delta\ct(g_0)$ and $\Delta\ctt(g_0)$ such that the difference between a fully
non-perturbative definition of $\ct$ and $\ctt$ and the perturbative 
estimates~(\ref{eq:ct}, \ref{eq:cttilde}) is likely to be covered.
We here choose 
\begin{equation}
  \Delta\ct(g_0) = \ct^{\rm eff} g_0^6,\qquad
  \Delta\ctt(g_0) = \ctt^{\rm eff} g_0^4,\qquad
\end{equation}
i.e.~a term of the neglected order with an effective coefficient. In the case of $\ct$ which is known to 2-loop order, 
cf.~Subsect.~\ref{subsec:ctctt}, we use a geometric progression and define
\begin{equation}
   \ct^{\rm eff} = \left(\ct^{(2)}/\ct^{(1)}\right) \times \ct^{(2)}  = 0.74104 \times \ct^{(2)} = -0.01734\,.
\end{equation}
For $\ctt$ we simply use
\begin{equation}
   \ctt^{\rm eff} = \ctt^{(1)} = -0.01795 \,.
\end{equation}
We note that particularly the choice for $\Delta\ct$ is likely an overestimate, due to the accidental cancellation
of the gluonic and fermionic terms observed in~Subsect.~\ref{subsec:ctctt}.

There are several options for the inclusion of this systematic error. We chose to proceed as
follows: we first perform continuum extrapolations ignoring potential O($a$) errors in both
the original and the shifted data. We then take the spread of a given observable as
an additional systematic error and add it in quadrature. Obviously this assumes that
this systematic error is subdominant. We have therefore dismissed all continuum extrapolations where 
this turned out not to be the case. We will discuss the impact of these variations on the 
continuum extrapolations in the next section.

\section{Continuum results}
\label{sec:cont}

\subsection{Continuum extrapolation of the step-scaling function}

\begin{figure}[!hptb]
     \centering
      \includegraphics[scale=1.0]{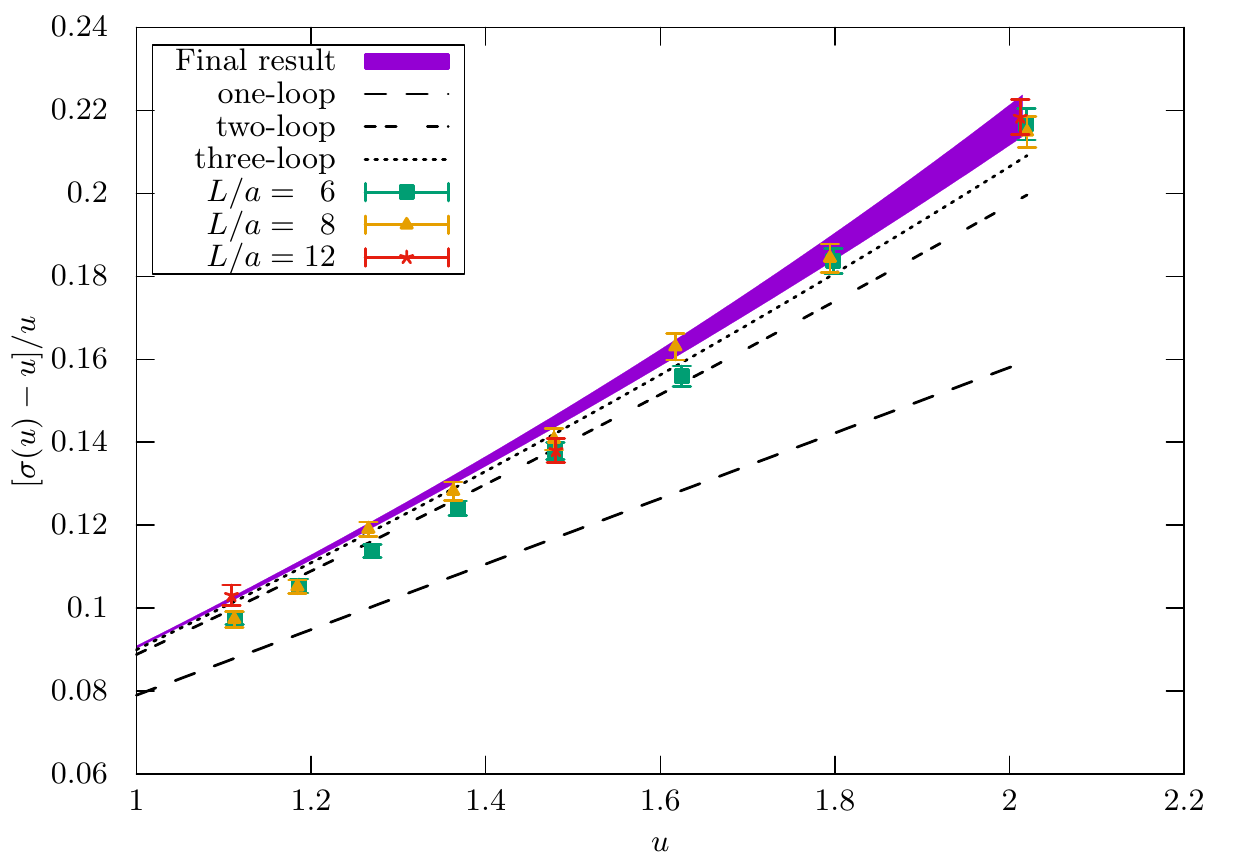}
      \caption{The step-scaling function for the $\nu=0$
        SF-coupling. The band shows our result (fit C,
        cf.~table~\ref{tab:Sigfits}). The data points are the
        approximations at finite $L/a=6,8,12$ 
     taken from table~\ref{tab:rawdata2} with errors from Eq.~(\ref{eq:ssf-error}).
     }
     \label{fig:ssf}
\end{figure}

We now proceed with the continuum extrapolation of the data for the step-scaling function,
for our default scheme with $\nu=0$. The 19 available data points for lattice resolutions $L/a=6,8,12$ are 
shown in figure~\ref{fig:ssf}.  Simulation parameters have been chosen 
such as to have approximately matched $u$-values between
different $L/a$, and this is seen in the vertical line-up of the data.
The fact that the data are so close together at given $u$-value illustrates that cutoff effects 
in the SF scheme with the chosen lattice regularization
are generally small, even without perturbative improvement.
While our data enables a more traditional continuum extrapolation, $u$-value by $u$-value, 
we have done this only as a cross-check. Our preferred strategy is to simultaneously 
fit all data to a global ansatz of the form
\begin{eqnarray}
    \Sigma^{(i)}(u,a/L) &=& \sigma(u) + \rho^{(i)}(u) \, (a/L)^2 \, .
    \label{eq:Sigmafit}
\end{eqnarray}
Here $i=1,2$ denotes the order of perturbative improvement of $\Sigma$ and $i=0$ 
refers to unimproved data. In general, such global fits have the advantage
that an interpolation of the data to common $u$-values is not required. More importantly,
however, the expected smooth $u$-dependence of the step-scaling function
both on the lattice and in the continuum limit, is automatically built into this ansatz.
As anticipated in the last section, we assume leading cutoff effects
to start at O($a^2$), with the linear $a$-effects being treated as
systematic errors. Our fit ans\"atze for the cutoff effects thus are of the form,
\begin{eqnarray}
    \rho^{(i)}(u) &=& \sum_{k=1}^{n_\rho^{}} \rho^{(i)}_k u^{i+1+k}\, ,
\end{eqnarray}
and the assumption of no lattice artefacts, $\rho^{(i)} = 0$,  is referred to by 
$n_\rho^{}=0$. For the continuum step scaling function
we consider polynomial fits with $n_c=2$ parameters,
\begin{equation}
  \sigma(u) = u + s_0 u^2 + s_1 u^3 + c_1 u^4 + c_2 u^5,
  \label{eq:2parfit}
\end{equation}
or 1-parameter fits  ($n_c=1$),
\begin{equation}
  \sigma(u) = u + s_0 u^2 + s_1 u^3 + s_2 u^4 + c_1 u^5,
  \label{eq:1parfit}
\end{equation}
where  $s_{0,1,2}$ are fixed to their perturbative
values~Eqs.~(\ref{eq:s012}). 
As the lattice artefacts are generally small at most $n_\rho=2$  parameters are required
to obtain excellent fits to the data. A selection of our fits is given in table~\ref{tab:Sigfits}.
As an example we consider a 4-parameter fit (fit D) with $n_c=n_\rho=2$ 
to the 2-loop improved data at $\nu=0$,
\begin{eqnarray}
         \Sigma^{(2)}(u,a/L) &=& u + s_0 u^2 + s_1 u^3 + c_1 u^4 + c_2 u^5 \nonumber \\
                              &&\; + (\rho_1 u^4 + \rho_2 u^5) (a/L)^2 \, .
\end{eqnarray}
Including all lattices with $L/a\ge 6$ there are thus 19 available data points
and 4 fit parameters in the 5th order polynomial in $u$. The fit has an excellent 
$\chi^2/{\rm d.o.f} = 14.5/15$ with the continuum parameters and their covariance
given by
\begin{equation}
   c_1 = 0.0014(3),\qquad 
   c_2 = 0.0005(2) ,\qquad    
   {\rm Cov}(c_1,c_2)= -0.38\times 10^{-5}.
  \label{eq:fitD}
\end{equation}
Note that the fit coefficient $c_1$ is not far from the perturbative
value $s_2=0.001151$; it is therefore reasonable to fix this parameter to
the perturbative one and only fit a next order coefficient. Hence
the majority of fits in table~\ref{tab:Sigfits} only have $n_c=1$ continuum parameters, either
$c_1$ in Eq.~(\ref{eq:1parfit}), or a 4-loop coefficient in the $\beta$-function, $b_3^\text{fit}$
(cf.~Subsection \ref{subsec:betaeff} below).


\newcommand{\dof}{\mathrm{dof}}

\begin{table}[h!]
  \small
  \centering
  \begin{tabular}{CcCCCCCCCCCC}\toprule
     \nu & fit  &  i  & \big[\tfrac{L}{a}\big]_\mathrm{min} 
                            & n_\rho^{} 
                                  & n_c & c_1\times 10^4                 & b_{3,\nu}^\mathrm{fit} \times (4\pi)^4 
                                                                                &
                                                                                  \chi^2 & \dof \\ 
    \midrule
    ~0   & A    &  0  &  6  &  2  &  1  & 6(2)(3)
                                                                         &      & 14.7   &  16  \\ 
    ~0   & B    &  1  &  6  &  2  &  1  & 5(3)(3)
                                                                         &      & 14.2   &  16  \\ 
    ~0   & B$'$ &  1  &  6  &  1  &  1  & 8(2)(2)
                                                                         &      & 18.4   &  17  \\ 
    ~0   & C    &  2  &  6  &  2  &  1  & 6(3)(3)
                                                                         &      & 14.5   &  16  \\ 
    ~0   & D    &  2  &  6  &  2  &  2  &
                                          \text{cf.~Eq.}~\eqref{eq:fitD} &      & 14.5   &  15  \\ 
    ~0   & E    &  2  &  6  &  2  &  1  &
                                                                      & 4(2)(2) & 14.6   &  16  \\ 
    ~0   & F    &  2  &  8  &  1  &  1  &
                                                                      & 4(3)(3) & 12.7   &  ~9  \\ 
    ~0   & G    &  2  &  8  &  0  &  2  &
                                          \text{cf.~Eq.}~\eqref{eq:fitG} &      & 13.0   &  ~9  \\ 
    ~0   & H    &  1  &  6  &  2  &  1  &
                                                                      & 3(2)(3) & 14.1   &  16  \\ 
    \midrule
    ~0.3 & A    &  0  &  6  &  2  &  1  & 3(2)(3)
                                                                         &      & 21.2   &  16  \\ 
    ~0.3 & B    &  1  &  6  &  2  &  1  & 1(2)(3)
                                                                         &      & 20.0   &  16  \\ 
    ~0.3 & B$'$ &  1  &  6  &  1  &  1  & 3(2)(2)
                                                                         &      & 20.8   &  17  \\ 
    ~0.3 & H    &  1  &  6  &  2  &  1  &
                                                                      &~0(2)(2) & 20.0   &  16  \\ 
    \midrule
    -0.5 & A    &  0  &  6  &  2  &  1  & 12(5)(5)
                                                                         &      & 11.6   &  16  \\ 
    -0.5 & B    &  1  &  6  &  2  &  1  & 15(5)(5)
                                                                         &      & 10.4   &  16  \\ 
    -0.5 & B$'$ &  1  &  6  &  1  &  1  & 24(4)(4)
                                                                         &      & 18.4   &  17  \\ 
    -0.5 & H    &  1  &  6  &  2  &  1  &
                                                                      &11(5)(5) & 10.4   &  16  \\ 
    \bottomrule
  \end{tabular}
  \caption{Overview of the continuum fit functions and results. The naming
           convention is the same as in ref.~\cite{Brida:2016flw}. The two errors
           in the fit parameters are the statistical and
           the total error respectively, where the total error includes the
           systematic uncertainty from a variation of $\ct$ and $\ctt$, added in
           quadrature.
          }
  \label{tab:Sigfits}
\end{table}

Given the smallness of the cutoff effects, even  fit G with $n_\rho=0$ parameters seems reasonable, 
if one restricts to data with $L/a\ge 8$. For the 2 continuum fit parameters of fit G the results are,
\begin{equation}
   c_1 = 0.0006(12),\qquad 
   c_2 = 0.0011(7),\qquad    
   {\rm Cov}(c_1,c_2)= -0.86\times 10^{-6}\,.
   \label{eq:fitG}
\end{equation}
While the $\chi^2/{\rm d.o.f} = 13/9 =1.44$ does not look too good,
a comparison with fits B$'$ and $F$ (with $n_\rho=1$)
indicates that this may be an accident. In fact the $\chi^2$-values are not a
sharp criterion in our case, as these strictly refer only to the statistical errors 
of the data and the given fit functions used, and thus do not account 
for the systematic uncertainties from cutoff effects linear in $a$.

In order to quantify these systematic uncertainties we repeat 
the fits with the data shifted by varying either $\ct$ or $\ctt$,
as explained in Subsect.~\ref{subsec:oamodel}. 
For fits with a single continuum parameter, $n_c=1$, we then take
the spread in central values for this parameter as a systematic uncertainties
due to either $\ct$ or $\ctt$ variations and combine them 
in quadrature with the statistical error to obtain a total error of the fit parameter.
Thus, in table~\ref{tab:Sigfits}, the fits with $n_c=1$ show 2 errors, the first being
the statistical and the second the total error. In all fits
we find that the $\ct$-uncertainty dominates the effect of the $\ctt$-uncertainty; 
for instance, for fit B we obtain
\begin{equation}
  c_1 = [49(25)_{\text{stat.}}(15)_{\Delta \ct}(6)_{\Delta \ctt}]\times 10^{-5} = 
  5(3)_\text{stat.}(3)_\text{total}\times 10^{-4}\,, 
\end{equation}
where the r.h.s.~takes the form given in table~\ref{tab:Sigfits}.
For fits with $n_c=2$ continuum parameters we proceed in the same way. However, 
rather than quoting a total error on the continuum fit parameters, we propagate these
uncertainties to the observables in table~\ref{tab:coupling-Lambda-b3eff},
where the results from the $n_c=2$ fits D and G are given with both a statistical
and total error.

While the total errors for most fits are dominated by the statistical error,
this is not the case of fit G, where the total errors are predominantly systematic,
cf. table~\ref{tab:coupling-Lambda-b3eff}. This indicates that fits with $n_\rho=0$ are too rigid
to account for the O($a$) variation of the data. While $n_\rho=1$ fits B$'$ and F 
are acceptable, we settled for fit ans\"atze with $n_\rho=2$  and $n_c=1$ to data 
with $L/a\ge 6$ as our preferred choice (fits A,B,C,E,H).
Then, using the 2-loop improved data leaves us with fits C and E, which are essentially
equivalent, and figure~\ref{fig:ssf} shows $\sigma(u)$ from fit~C with its error band.

\subsection{The SF coupling for $\nu=0$  at scales $L_n=L_0/2^n$}

We now use the continuum fit functions for the step-scaling function at $\nu=0$ to evaluate the
coupling at different scales $L_n = L_0/2^n$, separated by factors of 2.
Our starting point is the reference scale $L_0$, 
defined implicitly by
\begin{equation}
  \gbar^2(L_0) = 2.012\,.
  \label{eq:Lzero}
\end{equation}
The value $2.012$ corresponds to the largest value of the coupling $u$ for which
the step-scaling function is known. In physical units the scale $L_0$ has been determined 
to be around $1/(4\,{\rm GeV})$ \cite{Bruno:2017gxd}.
We note that $\sigma(2.012)$  defines the coupling $\gbar^2(2L_0)$, 
so that the lowest energy scale reached with the SF coupling is around 2~GeV.

Recursive application of the continuum step scaling function, $\sigma(u)$, 
allows us to obtain, in the continuum limit, 
the couplings at $L_n= L_0/2^n$, where $n=-1,0,1,2,\ldots$, 
via\footnote{The recursion towards larger $n$ requires a numerical inversion of the step-scaling function. 
This is not a problem given that the step-scaling function is, in practice, found to be 
a monotonously increasing function for the range of couplings considered here.}
\begin{equation}
    u_n = \sigma(u_{n+1}),  \qquad u_n = \gbar^2(L_n).
\end{equation}
This defines the couplings $u_n$ as a set of observables, with our data
enabling the recursion up to $n=5$, thereby  covering a total scale factor of $L_{-1}/L_5 = 2^6=64$.  
The results for $u_n$ are collected in table~\ref{tab:coupling-Lambda-b3eff}, for the various
fit functions representing $\sigma(u)$.
\begin{table}[p]
  \small
  \centering
  \vskip-5em
  \begin{tabular}{cllllll}\toprule
   $n$& \multicolumn{6}{c}{$\nu = 0$}                                                 \\\cmidrule(lr){2-7}
      & fit A      & fit B     & fit B$'$  & fit C      & fit D        & fit G        \\\cmidrule(lr){2-7}
      & $u_n$:                                                                        \\[0.2em]
    0 &  2.012     & 2.012     & 2.012     &  2.012     & 2.012        & 2.012        \\ 
    1 &  1.712(3)  & 1.714(3)  & 1.710(3)  &  1.712(3)  & 1.712(3)(3)  & 1.711(1)(5)  \\ 
    2 &  1.493(4)  & 1.495(4)  & 1.490(3)  &  1.493(4)  & 1.493(4)(5)  & 1.492(2)(7)  \\ 
    3 &  1.326(4)  & 1.327(4)  & 1.322(3)  &  1.325(4)  & 1.325(5)(6)  & 1.324(2)(8)  \\ 
    4 &  1.193(4)  & 1.194(4)  & 1.190(3)  &  1.193(4)  & 1.192(5)(6)  & 1.191(2)(8)  \\ 
    5 &  1.085(3)  & 1.086(4)  & 1.082(3)  &  1.085(4)  & 1.084(5)(6)  & 1.084(3)(8)  \\ 
 $-1$ &  2.450(10) & 2.447(10) & 2.458(8)  &  2.451(10) & 2.451(10)(11)& 2.457(5)(12) \\\cmidrule(lr){2-7}  
      & $L_0 \Lambda \times 10^2$:                                                    \\[0.2em] 
    0 &  3.14      & 3.14      & 3.14      &  3.14      & 3.14         & 3.14         \\ 
    1 &  3.10(3)   & 3.11(3)   & 3.08(2)   &  3.10(3)   & 3.10(2)(3)   & 3.09(1)(4)   \\ 
    2 &  3.07(4)   & 3.09(5)   & 3.04(4)   &  3.07(5)   & 3.07(5)(6)   & 3.05(2)(8)   \\ 
    3 &  3.05(6)   & 3.08(6)   & 3.01(5)   &  3.05(6)   & 3.05(7)(8)   & 3.03(3)(11)  \\ 
    4 &  3.04(7)   & 3.06(7)   & 2.98(5)   &  3.03(7)   & 3.03(9)(11)  & 3.02(4)(14)  \\ 
    5 &  3.03(7)   & 3.06(8)   & 2.97(6)   &  3.02(8)   & 3.01(12)(14) & 3.00(5)(17)  \\\cmidrule(lr){2-7}  
      & $b_3^{\rm eff}\times (4\pi)^4$:                                               \\[0.2em]  
    0 &  3(2)      & 2(2)      & 4(1)      &  3(2)      &  2(5)(5)     &  5(2)(2)     \\ 
    1 &  3(2)      & 2(2)      & 5(2)      &  3(2)      &  3(5)(5)     &  5(2)(2)     \\ 
    2 &  4(3)      & 3(3)      & 6(2)      &  4(3)      &  4(4)(4)     &  6(2)(3)     \\ 
    3 &  4(3)      & 3(3)      & 7(2)      &  4(3)      &  4(3)(4)     &  6(2)(3)     \\ 
    4 &  5(3)      & 3(3)      & 7(3)      &  5(3)      &  5(3)(3)     &  7(1)(4)     \\ 
    5 &  5(3)      & 4(3)      & 8(3)      &  5(3)      &  5(3)(4)     &  7(1)(5)     \\\midrule
  $n$ & \multicolumn{3}{c}{$\nu = 0.3$}    & \multicolumn{3}{c}{$\nu =-0.5$}          \\\cmidrule(lr){2-4}\cmidrule(lr){5-7}
      & fit A      & fit B     & fit B$'$  & fit A     & fit B         & fit B$'$     \\\cmidrule(lr){2-4}\cmidrule(lr){5-7}  
      & $u_n$:     &           &           & $u_n$:                                   \\[0.2em]
    0 &   2.169    & 2.169     & 2.169     & 1.795     & 1.795         & 1.795        \\ 
    1 &   1.828(4) & 1.832(4)  & 1.829(3)  & 1.550(3)  & 1.548(4)      & 1.542(3)     \\ 
    2 &   1.582(5) & 1.587(5)  & 1.584(4)  & 1.366(5)  & 1.363(5)      & 1.356(4)     \\ 
    3 &   1.396(5) & 1.401(5)  & 1.398(4)  & 1.223(5)  & 1.220(5)      & 1.212(4)     \\ 
    4 &   1.250(4) & 1.255(4)  & 1.252(3)  & 1.108(5)  & 1.106(5)      & 1.098(4)     \\ 
    5 &   1.133(4) & 1.136(4)  & 1.134(3)  & 1.014(4)  & 1.012(4)      & 1.004(3)     \\ 
   $-1$ & 2.677(12)& 2.665(13) & 2.672(10) & 2.145(10) & 2.151(10)     & 2.168(8)     \\ \cmidrule(lr){2-4}\cmidrule(lr){5-7}  
      & $L_0 \Lambda \times 10^2$: & &     & $L_0 \Lambda \times 10^2$:               \\[0.2em]
    0 &   3.05     & 3.05      & 3.05      & 3.34      & 3.34          & 3.34         \\ 
    1 &   3.02(3)  & 3.05(3)   & 3.03(2)   & 3.28(4)   & 3.25(4)       & 3.19(3)      \\ 
    2 &   3.00(4)  & 3.05(5)   & 3.02(4)   & 3.23(7)   & 3.19(7)       & 3.09(5)      \\ 
    3 &   2.99(6)  & 3.04(6)   & 3.01(5)   & 3.20(8)   & 3.15(8)       & 3.01(6)      \\ 
    4 &   2.98(7)  & 3.04(7)   & 3.00(5)   & 3.17(10)  & 3.12(10)      & 2.96(7)      \\ 
    5 &   2.97(7)  & 3.04(8)   & 3.00(6)   & 3.15(11)  & 3.09(11)      & 2.91(8)      \\\cmidrule(lr){2-4}\cmidrule(lr){5-7}  
      & $b_3^{\rm eff}\times (4\pi)^4$:& & &  $b_3^{\rm eff}\times (4\pi)^4$:         \\[0.2em]
    0 &   2(2)     & 0(2)      & 1(1)      &  5(4)     &  7(4)         & 13(3)        \\ 
    1 &   2(2)     & 0(2)      & 1(2)      &  7(4)     &  9(4)         & 17(3)        \\ 
    2 &   2(2)     & 0(2)      & 1(2)      &  8(5)     & 11(5)         & 19(4)        \\ 
    3 &   3(2)     & 0(3)      & 2(2)      &  9(5)     & 12(6)         & 21(4)        \\ 
    4 &   3(3)     & 0(3)      & 2(2)      &  9(6)     & 13(6)         & 23(5)        \\ 
    5 &   3(3)     & 0(3)      & 2(2)      & 10(6)     & 13(6)         & 24(5)        \\ 
   \bottomrule
  \end{tabular}
  \caption{Results for the couplings $u_n=\gbar^2_\nu(L_n)$, the $\Lambda$-parameter evaluated at $u_n$, cf.~Eq.~(\ref{eq:L0Lambda}), 
  in units of the reference scale, $L_0$~(\ref{eq:Lzero}), and the effective $\beta$-function coefficient, $b_3^{\rm eff}$~(\ref{eq:b3eff}), 
  for most fits of table~\ref{tab:Sigfits}. Results for $L_0\Lambda$ obtained
  with fits E, F and H are given in table~\ref{tab:L0Lambda_b3fit}.}
    \label{tab:coupling-Lambda-b3eff}
\end{table}

\subsection{Effective and fitted $\beta$-function}
\label{subsec:betaeff}

Given $\sigma(u)$  in terms of 1 or 2 continuum parameters $c_k$, 
one may translate this result into an effective 3-loop coefficient of the continuum $\beta$-function.
For convenience we define $b(g^2) = -g\beta(g)$ so that
\begin{equation}
  b(u) = b_{\rm 3loop}(u) + b_3^{\rm eff} u^5, \qquad  b_{\rm 3loop}(u)= b_0 u^2 + b_1 u^3+ b_2 u^4.
\end{equation}
Then Eq.~(\ref{eq:beta_sigma}) becomes,
\begin{equation}
   \int_{u}^{\sigma(u)} \frac{{\rmd} v}{b(v)} = 2\ln 2\,.
  \label{eq:b_sigma}
\end{equation}
Differentiation w.r.t.~$u$ yields
\begin{equation}
  \frac{\sigma'(u)}{b(\sigma(u))} -\frac{1}{b(u)} =0\,, 
\end{equation}
which can be solved for $b_3^{\rm eff}$, with the result,
\begin{equation}
   b_3^{\rm eff} = \frac{b_{\rm 3loop}(u)\sigma'(u)-b_{\rm 3loop}(\sigma(u))} { \sigma^5(u)-u^5\sigma'(u)}\,.
   \label{eq:b3eff}
\end{equation}
Note that $b_3^{\rm eff}$ will depend on the value $u$ where it is measured. Extracting this coefficient at different 
values of $u$ should yield consistent results in the perturbative regime, and this is
indeed the case for the $\nu=0$ data, cf.~table~\ref{tab:coupling-Lambda-b3eff}.

This motivates a different parameterization of our fits with a single continuum parameter, namely
via a 4-loop coefficient $b^{\rm fit}_3$ in the $\beta$-function as a 
fit parameter\footnote{In ref.~\cite{Brida:2016flw} this fit parameter was denoted $b_3^{\rm eff}$.}.
This is the purpose of fits E, F and H, cf.~table~\ref{tab:Sigfits}, where
we have taken $\sigma(u)$ to be defined by Eq.~(\ref{eq:b_sigma}) with 
$b(u) = b_{\rm 3loop}(u) + b_3^{\rm fit} u^5$
and inserted $\sigma(u)$ into Eq.~(\ref{eq:Sigmafit}). 
The resulting values for the fit parameter $b_3^{\rm fit}$ are given in table~\ref{tab:Sigfits}. 
This representation of our continuum results is very practical. While the fit function in Eq.~(\ref{eq:1parfit})
allows us to find the couplings at scales which are separated by a factor 2, the
$\beta$-function readily yields the scale ratio separating two given couplings.

\subsection{Determination of the $\Lambda$-parameter} \label{subsec:L0Lambda}

Once the coupling $u_n = \gbar^2(L_n)$ is small enough, it is justified to use
three-loop perturbation theory for the $\beta$-function in the expression
\begin{eqnarray}
 L_0 \Lambda &=& 2^{n} \left( b_0 \gbar^2(L_n) \right)^{-b_1/(2b_0^2)} e^{-1/(2b_0 \gbar^2(L_n))} \nonumber \\
    && \times \exp\left\{-\int\limits_0^{\gbar(L_n)}{\rm d}x\ \left[\frac{1}{\beta(x)} 
        +\frac{1}{b_0x^3} - \frac{b_1}{b_0^2x} \right] \right\} \, ,
    \label{eq:L0Lambda}
\end{eqnarray}
and determine the $\Lambda$-parameter in units of $L_n$ and thus in units of $L_0 = 2^n L_n$.
Note that the expansion of the integral in the exponent
\begin{equation}
  \int\limits_0^{\gbar}{\rm d}x\ \left[\frac{1}{\beta(x)} 
                +\frac{1}{b_0x^3} - \frac{b_1}{b_0^2x} 
                               \right] =  
                               \frac{b_0b_2-b_1^2} {2b_0^3} \gbar^2 +  
                               \frac{b_0^2 b_3-2 b_0b_1b_2+b_1^3} {4b_0^4} \gbar^4 + \rmO(\gbar^6)\,,
\label{eq:Lambda-expand}
\end{equation}
is unknown at order $\gbar^4$ as this term 
requires the knowledge of the 4-loop coefficient $b_3$ which is not available in the SF scheme.
Provided such higher order terms are small, the result for $L_0\Lambda$ should be independent of $n$ 
and the way the integral is evaluated. For completeness we note that our default evaluation consists 
in the direct numerical integration, using the truncated 3-loop $\beta$-function
without expansion of the integrand or the exponential function.
%
The results for $\Lambda$ in units of $L_0$ are given in table~\ref{tab:coupling-Lambda-b3eff}, where
Eq.~(\ref{eq:L0Lambda}) is evaluated for the coupling at scales $L_n$, for $n=0,\ldots,5$ and for the various fit functions.

An alternative evaluation of the $\Lambda$-parameter is obtained with the fits E, F and H in terms
of a fitted $\beta$-function. One simply inserts the $\beta$-function into
Eq.~(\ref{eq:L0Lambda}) and evaluates the integral numerically between $\gbar^2(L_0)=2.012$ and $\gbar^2(0)=0$.
The resulting $\Lambda$-parameters are given in table~\ref{tab:L0Lambda_b3fit} and show a remarkable consistency.
We will discuss the results further in Subsect.~\ref{subsec:Lambdanu}.

\begin{table}[h]
\centering
  \begin{tabular}{LCCL}\toprule
            & \multicolumn{3}{C}{L_0\Lambda \times 10^2}  \\ 
   \multicolumn{1}{C}{\nu} & \text{fit E} & \text{fit F}  & \text{fit H} \\  \midrule
   \hphantom{+} 0        &  3.00(8)     & 3.01(10)       & 3.04(9)   \\    
   -0.5                  &              &                & 3.03(14)\\
   \hphantom{+}0.3       &              &                & 3.04(8)\\
   \bottomrule
  \end{tabular}
   \caption{$L_0\Lambda$ obtained with the fits to the 
   coefficient $b_3^\mathrm{fit}$ in the $\beta$-function, cf.~table~\ref{tab:Sigfits} 
   and Subsect.~\ref{subsec:L0Lambda}.}
\label{tab:L0Lambda_b3fit}
\end{table}

\subsection{Continuum extrapolation of $\Omega$ and $\tilde{\Omega}$}

The continuum extrapolation for $\Omega(u,a/L)$ and $\tilde\Omega(u,a/L)$ proceeds along the
same line as for the step-scaling function. A difference is that
both data sets can be constrained to the same continuum limit but require separate fit
coefficients for the cutoff effects. Moreover, the lattice resolutions $L/a$ cover the 
range $6-24$, i.e.~a factor of 4 in scale and thus allow for an excellent control of the continuum limit.

\begin{table}[t]
 \small
 \centering
 \begin{tabular}{cCCLLLL}\toprule  
  fit & \{\Omega^{(i)}, \tilde{\Omega}^{(i)}\}^{\vphantom{1}} 
                           & \chi^2/\dof 
                                     & \omega(1.11) & \omega(1.5) & \omega(2.012) & \omega(2.45) \\ \midrule
    A & \!i\!=\!1, L/a\ge 6\! & 47.8/45 & 0.1368(8)(9)   & 0.1307(7)(8)  & 0.1201(8)(9)    & 0.1123(13)(13) \\ 
    A & \!i\!=\!1, L/a\ge 8\! & 33.5/37 & 0.1385(10)(10) & 0.1319(8)(9)  & 0.1199(9)(10)   & 0.1117(13)(13) \\ 
    A & \!i\!=\!0, L/a\ge 6\! & \bf 61.3/45 & 0.1350(9)(9)   & 0.1290(7)(8)  & 0.1193(9)(10)   & 0.1118(12)(12) \\
    A & \!i\!=\!0, L/a\ge 8\! & 33.5/37 & 0.1379(11)(11) & 0.1311(8)(9)  & 0.1193(10)(10)  & 0.1115(12)(13) \\ \midrule
    B & \!i\!=\!1, L/a\ge 6\! & 47.8/44 & 0.1368(10)(11) & 0.1307(7)(8)  & 0.1201(9)(9)    & 0.1123(13)(13) \\
    B & \!i\!=\!1, L/a\ge 8\! & 33.5/36 & 0.1385(12)(13) & 0.1319(9)(9)  & 0.1199(10)(10)  & 0.1117(13)(13) \\
    B & \!i\!=\!0, L/a\ge 6\! & \bf 60.6/44 & 0.1344(12)(12) & 0.1291(7)(8)  & 0.1192(9)(10)   & 0.1120(12)(13) \\
    B & \!i\!=\!0, L/a\ge 8\! & 33.5/36 & 0.1381(15)(15) & 0.1311(8)(9)  & 0.1194(10)(10)  & 0.1115(13)(13) \\
    \bottomrule
 \end{tabular}
 \caption{Results of the combined fits A and B for $\Omega^{(i)}(u,a/L)$ and
          $\tilde{\Omega}^{(i)}(u,a/L)$ with ($i=1$) and without ($i=0$)
          improvement. The 2 errors given are the statistical and the total
          error, respectively, where the latter includes an estimate of the
          remnant uncertaintly due to linear $a$-effects.
         }
  \label{tab:omega}
\end{table}

The global fit ans\"atze used here are
\begin{equation}
  \Omega^{(i)}(u,a/L) = \omega(u) + \rho^{(i)}(u,a/L)\,,
\end{equation}
and analogously for $\tilde\Omega^{(i)}$ with $\tilde\rho^{(i)}$. Here, $i=1,0$  
refers to 1-loop improved data (cf.~Subsect.~\ref{subsec:pertimp}) 
or unimproved data, respectively. In the models for the cutoff effects
we just include 2 quadratic terms in $a/L$ for either data set, with coefficients  $\rho_{1,2}$  
and $\tilde{\rho}_{1,2}$, e.g.
\begin{equation}
  \rho^{(i)}(u,a/L) = \left(\rho_1 u^i + \rho_2 u^{i+1}\right)\frac{a^2}{L^2}\,,
\end{equation}
and the powers of $u$ are chosen  according to the expectation from perturbation theory.
As in the case of the step-scaling function, linear terms in $a/L$ will be treated as systematic errors. 

The continuum function $\omega(u)$ is parameterized 
by a fourth order polynomial in $u$,
\begin{equation}
 \omega(u) = \begin{cases}  v_1+ v_2 u + \sum_{k=1}^3 d_k u^{k+1}\,, & \text{fits type A}, \\
                            v_1+ \sum_{k=1}^4 d_k u^{k}\,,           & \text{fits type B}\,,
  \end{cases}
  \label{eq:fitsAB}
\end{equation}
with fit parameters $d_k$, $k=1,\ldots,4$ and $v_1$ and $v_2$ set to the known perturbative
coefficients, Eq.~(\ref{eq:v_12}). We have also experimented with separate 
fits to $\Omega^{(i)}$ and $\tilde{\Omega}^{(i)}$
and find good overall consistency. 
Here, we restrict the discussion to combined fits of the $\Omega^{(i)}$ and $\tilde\Omega^{(i)}$
data, with a common continuum fit function, $\omega(u)$. 
We distinguish fits of type A and B with 3 and 4 continuum
fit parameters, respectively. Hence, fits of type A have  $3+2\times 2 =7$  parameters, while type B fits have 8 parameters.

With these fit ans\"atze one obtains decent  $\chi^2/{\rm d.o.f.}$  values for the one-loop improved data,
even when including all 52 data points with $L/a\ge 6$ (cf.~tables~\ref{tab:rawdata1} and \ref{tab:rawdata2}).
Given this much data we may afford  to exclude the $L/a=6$ lattices, thereby reducing the number of data points to 44 .
An example for the continuum function $\omega(u)$ thus obtained is
\begin{equation}
  \omega(u)\vert_{\text{fit A}, i=1, L/a\ge 8} = 0.14307 - 0.004693 u + 0.01284 u^2 -0.01480 u^3 + 0.003349 u^4.
\end{equation}
The fit has a $\chi^2/{\rm d.o.f.} = 33.5/37 $ and the covariance matrix for the fit parameters is given by
\begin{equation}
{\rm Cov}(d_i,d_j) = \begin{pmatrix}   1.286  &  -1.244  &  0.2922\\
                                      -1.244  &   1.231  & -0.2945\\
                                       0.2922 &  -0.2945 &  7.153
                     \end{pmatrix}\times 10^{-5}\,.
                     \label{eq:covdd}
\end{equation}
Note that the error encoded in the covariance matrix is only the statistical error. 
To account for the systematic effect estimated from the variation 
of the O($a$) counterterm coefficients $\ct$ and $\ctt$ (cf.~Subsect.~\ref{subsec:oamodel}),
we here proceed in complete analogy with the analysis of the step-scaling function.
In table~\ref{tab:omega} we quote 2 errors, the first statistical, the second including the
effect of a $\ct$ and $\ctt$-variation. This only marginally increases the errors, as is evident from
table~\ref{tab:omega}.
\begin{figure}[!hptb]
     \centering
      \includegraphics[scale=0.8]{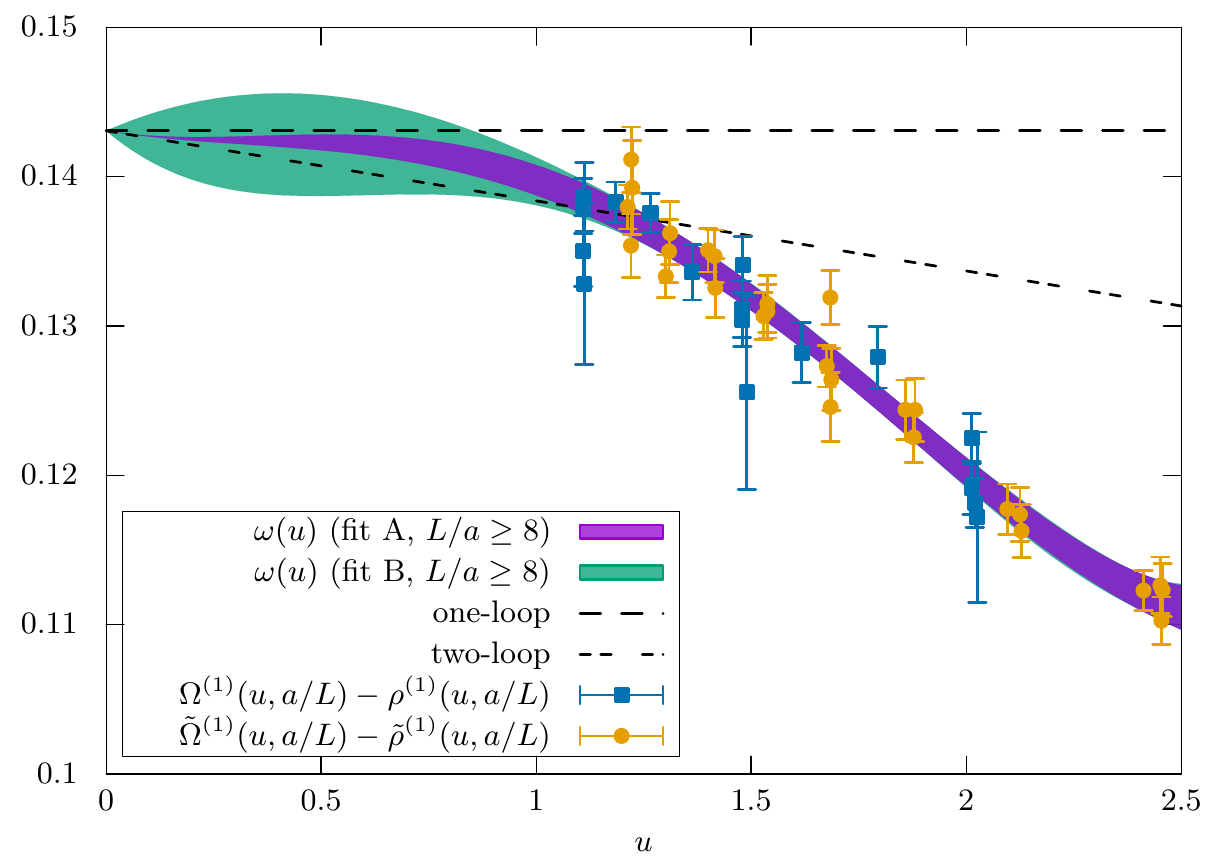}
       \caption{The bands show the continuum fit functions for fits  of type A and B 
       to one-loop improved data for $L/a\ge 8$ and the data points
       are one-loop improved data with the cutoff effects subtracted using the models $\rho^{(1)}(u,a/L)$ 
       and $\tilde\rho^{(1)}(u,a/L)$ from the type A fit.}
\label{fig:omega}
\end{figure}
The fits to the unimproved data have higher $\chi^2/{\rm d.o.f.}$ values,
emphasized in bold face in table~\ref{tab:omega}, unless the $L/a=6$ data
are dropped. As mentioned above, $\chi^2$ is not the full story, given
that our fits assume the absence of $a/L$ effects and this effect is
taken into account afterwards by our $\ct,\ctt$-variation. However,
we do see that 1) these variations have a tiny effect
on the continuum values and 2) still, for example, 
$\omega(1.11)$ of the large $\chi^2$ fits is off significantly. 
These fits have to be discarded. The other ones, which cover a remarkable range of lattice
spacings, are entirely consistent. 

These observations allow us to conclude that perturbative improvement works very well in our coupling range, 
our treatment of $\ct,\ctt$-variations is safe (maybe overly conservative),
and most importantly, resolutions $a/L \leq 1/6$ are sufficient 
to apply our continuum extrapolations which assume that $\rmO((a/L)^3)$
effects have a negligible effect. All this makes us very confident also in
the continuum extrapolations of $\Sigma$, where the very small
lattice spacings are not available, but where we have 2-loop perturbative
improvement at our disposal.

We return to the specific discussion of $\omega$. As our best value 
at the reference coupling $u=2.012 = \gbar^2(L_0)$ we choose
the result of fit A to 1-loop improved data with $L/a \ge 8$.
\begin{equation}
  \omega(2.012) = 0.1199(10),
  \label{eq:omref}
\end{equation}
which is required to define the starting point for the step-scaling procedure for SF$_\nu$ schemes with non-zero values of 
$\nu$ (s.~below). Another interesting value is $\omega(u)$ at the largest available coupling, $u = 2.45$,  which correspond to $\alpha=0.195$,
\begin{equation}
    \omega(2.45) = 0.1117(13),
\end{equation}
using the same fit function.
As discussed further in ~\cite{Brida:2016flw}, and as is evident from 
the large difference between 2-loop PT and
the non-perturbative result in Figure~\ref{fig:omega}, an unnaturally large 
next order perturbative coefficient would be required to 
perturbatively describe the function $\omega(u)$ at such values of the coupling.

Finally, we comment on the different behaviour of the fits A and B, which is seen in figure~\ref{fig:omega}
for small couplings, outside the range of the data. This illustrates the danger of using fit
functions outside their range of validity. While fit A is constrained to produce the 2-loop perturbative result 
for $\omega(u)$, Eq.~(\ref{eq:omega_pt}), fit B leaves the 2-loop coefficient $v_2$ as a fit parameter, $d_1$ (\ref{eq:fitsAB}). 
The result,
 \begin{equation}
    (4\pi)^2 d_1 = -0.9(2.9),
 \end{equation}
should be compared with Eq.~(\ref{eq:v_12}). While the central value is not too far off, 
the large error illustrates the difficulty to estimate such asymptotic coefficients, even if precise data
is available over a wide range of couplings.

\subsection{The step-scaling function for $\nu\ne 0$ and tests of perturbation theory} 
\label{subsec:Lambdanu}

\begin{figure}[!hptb]
     \centering
      \includegraphics[width=\textwidth]{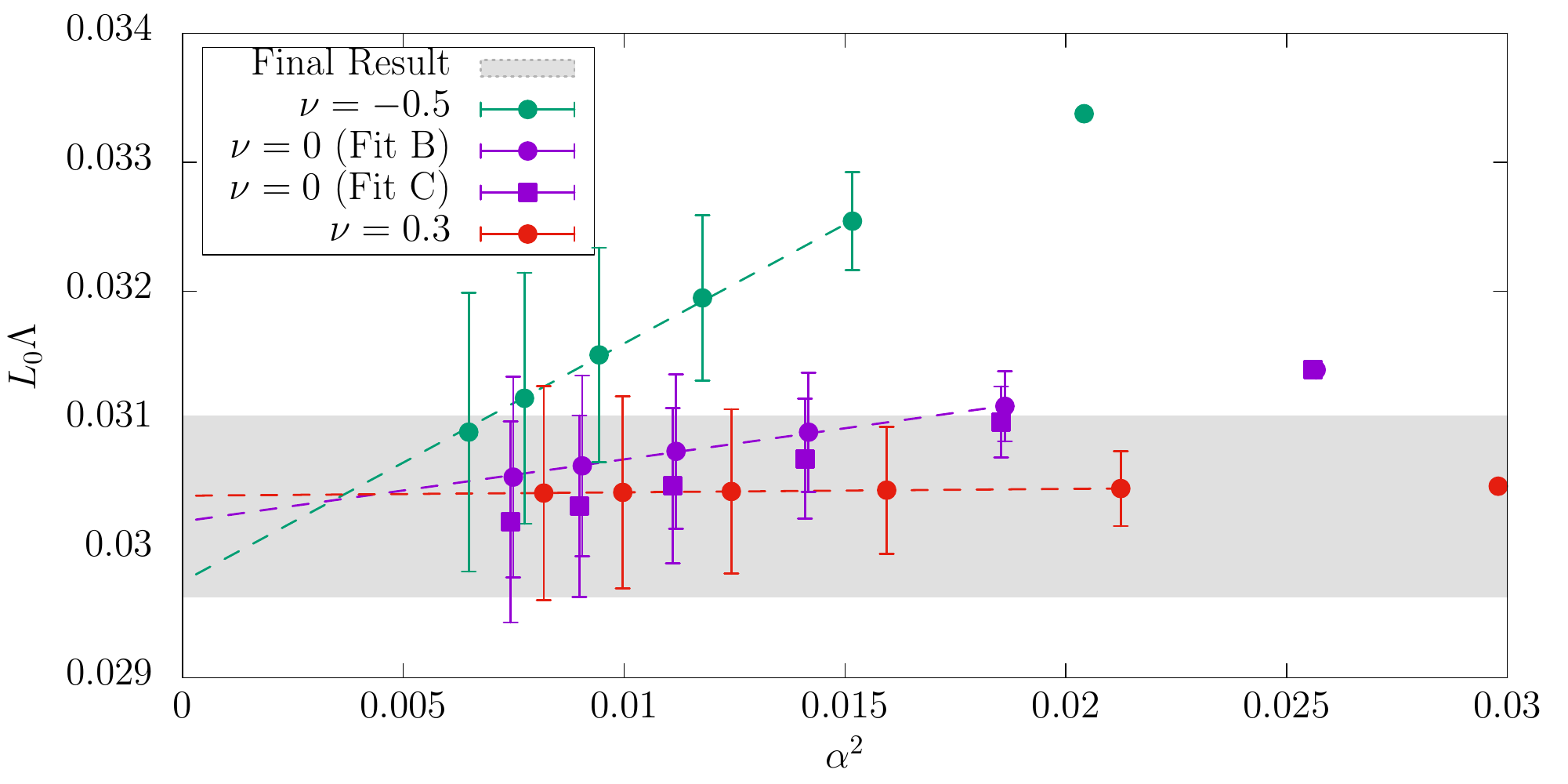}
      \caption{Determination of the $\Lambda$-parameter in units of $L_0$
      at different values of $\alpha$. We compare the
      extraction in different schemes ($\nu=-0.5,0,0.3$), and show a
      comparison with our final result
      Eq.~\eqref{eq:final_result}. As the reader can see when the
      extraction is performed at high enough energies ($\alpha\sim
      0.1$), all schemes nicely agree. See the text for a full discussion.}
     \label{fig:Lambda-vs-alphasq}
\end{figure}

The step-scaling functions for $\nu\ne 0$ can be treated in the same way as for $\nu=0$. 
The fit ans\"atze we have considered for $\Sigma_\nu(u,a/L)$ and 
$\Sigma_\nu^{(1)}(u,a/L)$ are analagous to the $\nu=0$ case, cf.~table~\ref{tab:Sigfits}.
Choosing the values $\nu=-0.5,0.3$ for illustration and fits of type A, B, B$'$ 
we quote again the couplings at $L_n$, as well as the results for $L_0\Lambda$.  
Here, the $\Lambda$-parameter is again the one of the $\nu=0$ scheme, 
i.e.~we use the known ratio of $\Lambda$-parameters (\ref{eq:Lambda_Lambdanu})
and evaluate, at $u_n^{}= \gbar^2_\nu(L_n)$, the expression
\begin{eqnarray}
 L_0 \Lambda &=& 2^{n} \frac{\Lambda}{\Lambda_\nu}   \varphi_\nu^{}\left(\gbar^2_\nu(L_n)\right),\qquad n=0,\ldots,5.  
    \label{eq:L0Lambda_nu}
\end{eqnarray}
The step-scaling procedure for $\nu\ne 0$ requires $\gbar^2_\nu(L_0)$ as 
starting point, which is given by 
\begin{equation}
   \frac{1}{\gbar_\nu^2(L_0)} =   \frac{1}{\gbar^2(L_0)} - \nu\omega\left(\gbar^2(L_0)\right) 
   = \frac{1}{2.012} - \nu\times 0.1199(10) \,.
\end{equation}
Note that this start value  now has a small uncertainty, due to the fact that $L_0$ is
still defined by Eq.~(\ref{eq:Lzero}) and the connection requires 
the result for $\omega(2.012)$ from Eq.~(\ref{eq:omref}).
For our choices of $\nu$-values this uncertainty is a factor 2-3 below the 
statistical uncertainty, and will be neglected in the following.
The propagation of errors to the couplings at scales $L_0/2^n$ 
and to $L_0\Lambda$ then proceeds in the same way as for $\nu=0$. Results are given 
in table~\ref{tab:coupling-Lambda-b3eff},
and in figure~\ref{fig:Lambda-vs-alphasq}.
We observe a roughly linear behaviour in $\alpha^2$, which suggests to directly fit to an effective 
4-loop coefficient $b_3^{\rm fit}$ in the $\beta$-function. 
This is done in fits $E,F$ and $H$, cf.~table~\ref{tab:Sigfits}. Not surprisingly, the resulting
fit coefficients roughly agree with the effective 4-loop coefficients, Eq.~(\ref{eq:b3eff}), 
given in table~\ref{tab:coupling-Lambda-b3eff}.
We also note that schemes at different $\nu$-values behave quite differently; the 2 chosen non-zero values
of $\nu$ illustrate this: while $\nu=0.3$ data shows no significant remnant
$\alpha$-dependence even up to $\alpha\approx 0.2$, the slope in $\alpha^2$ is very pronounced for $\nu=-0.5$. 
Therefore, it is a strong consistency check for our analysis that 
all values for $L_0\Lambda$ are compatible around $\alpha=0.1$,
despite considerable deviations at larger couplings. This means we can confidently 
extract $L_0\Lambda$ in this regime. Our final value is obtained from fit C, taking the $n=4$ estimate at $\nu=0$, viz
\begin{equation}
  \label{eq:final_result}
  L_0\Lambda = 0.0303(7) \quad \Rightarrow L_0\Lambda_\MSbar^{\Nf=3} = 0.0791(19)\,,
\end{equation}
which is slightly more precise than the result quoted in \cite{Brida:2016flw}, due to 
a refined model for the O($a$) boundary effects, cf.~Subsect.~\ref{subsec:oamodel}.
For an even more conservative error estimate one could take fit D, 
again at $n=4$ and $\nu=0$, 
which yields $L_0\Lambda =0.0303(11)$.   

Using the fits E, F and H, in terms of the fitted $\beta$-function,
the values in table~\ref{tab:L0Lambda_b3fit} are obtained. The fact that these
are all compatible, with very similar central values further boosts
the confidence that our final result is very robust. Finally, coming back to
the question raised in Section~2 about exponentially suppressed contributions, we
emphasize that the consistency of our analysis with fits taking the same functional form as
higher order perturbative terms provides indirect evidence for the absence of 
such non-standard terms within our numerical precision.

\subsection{Alternative tests}

So far, our strategy has been to first determine $\Lambda$ in the SF scheme and
then convert it to $\Lambda_{\MSbar}$. However, one might also match the 
SF to the $\MSbar$-coupling at 2-loop order using Eq.~(\ref{eq:alphaMSbar-alphaSFnu}) 
and then extract the $\Lambda$-parameter within the $\MSbar$-scheme.
While the perturbative precision is parametrically the same
as before, we present this alternative  view here, as it is closer 
to the strategy often used in phenomenological
applications.

Within the $\MSbar$-scheme we have, with $\mu=s/L$, and $L_n=L_0/2^n$,
\begin{eqnarray}
  \Lambda_{\MSbar}L_0 &=& s \frac{L_0}{L_n} \varphi^{}_{\MSbar}\left(\gbar^{}_{\MSbar}(L_n/s)\right) 
  \nonumber \\
  &=& s\, 2^n  
  \varphi^{}_{\MSbar}\left(\sqrt{\gbar^2_\nu(L_n)+p^\nu_1(s) \gbar^4_\nu(L_n)+p^\nu_2(s)\gbar^6_\nu(L_n)
   + \rmO\left[\gbar_\nu^8(L_n)\right]}\right),
   \label{eq:Lambda-via-MSbar}
\end{eqnarray}
where $s$ is an additional scale parameter and $p_i^{\nu}(s) = c_i^\nu(s)/(4\pi)^i$, cf.~Eq.~(\ref{eq:alphaMSbar-alphaSFnu}).
The unknown 3-loop and higher order terms in 
the argument of $\varphi^{}_{\MSbar}$ will be neglected in the following.
The function $\varphi^{}_{\MSbar}$, Eq.~(\ref{e:phig})
can be evaluated using up to 5-loop order for the $\beta$-function. 
For our range  of $\alpha$-values, the numerical  difference between 4- or 5-loop order evaluation is found to be negligible. 
The dominant uncertainty is due to the 2-loop truncation 
of the perturbative conversion to the $\MSbar$ coupling,
\begin{equation}
  \Delta g^2_\MSbar = \rmO\left[\gbar_\nu^8(L_n)\right] = \rmO\left[\gbar_\MSbar^8(L_n/s)\right]\,,
\end{equation}
which multiplies the sensitivity to a change in the coupling,
\begin{equation}
   \dfrac{d}{d g^2} \varphi^{}_{\MSbar}\left(g\right) \propto \frac{1}{g\beta_\MSbar^{}(g)} = \text{O($g^{-4}$)}\,, 
\end{equation}
and thus induces an  O($g^4$)  or O($\alpha^2$)  uncertainty in the estimate of the $\Lambda$-parameter.
As mentioned before, this is parametrically the same as previously, cf.~Eq.~(\ref{eq:Lambda-expand}).
\begin{figure}[!hptb]
     \centering
      \includegraphics[width=0.45\textwidth]{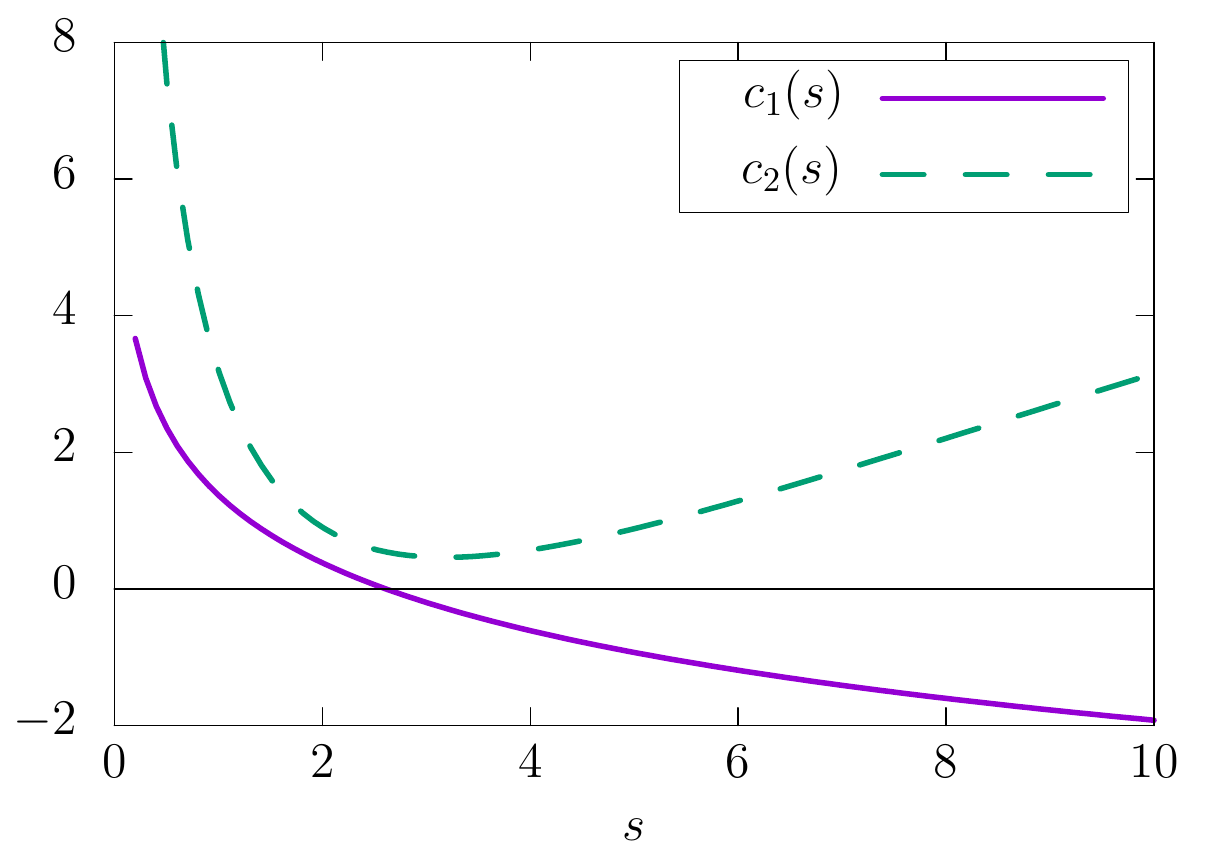}
      \includegraphics[width=0.45\textwidth]{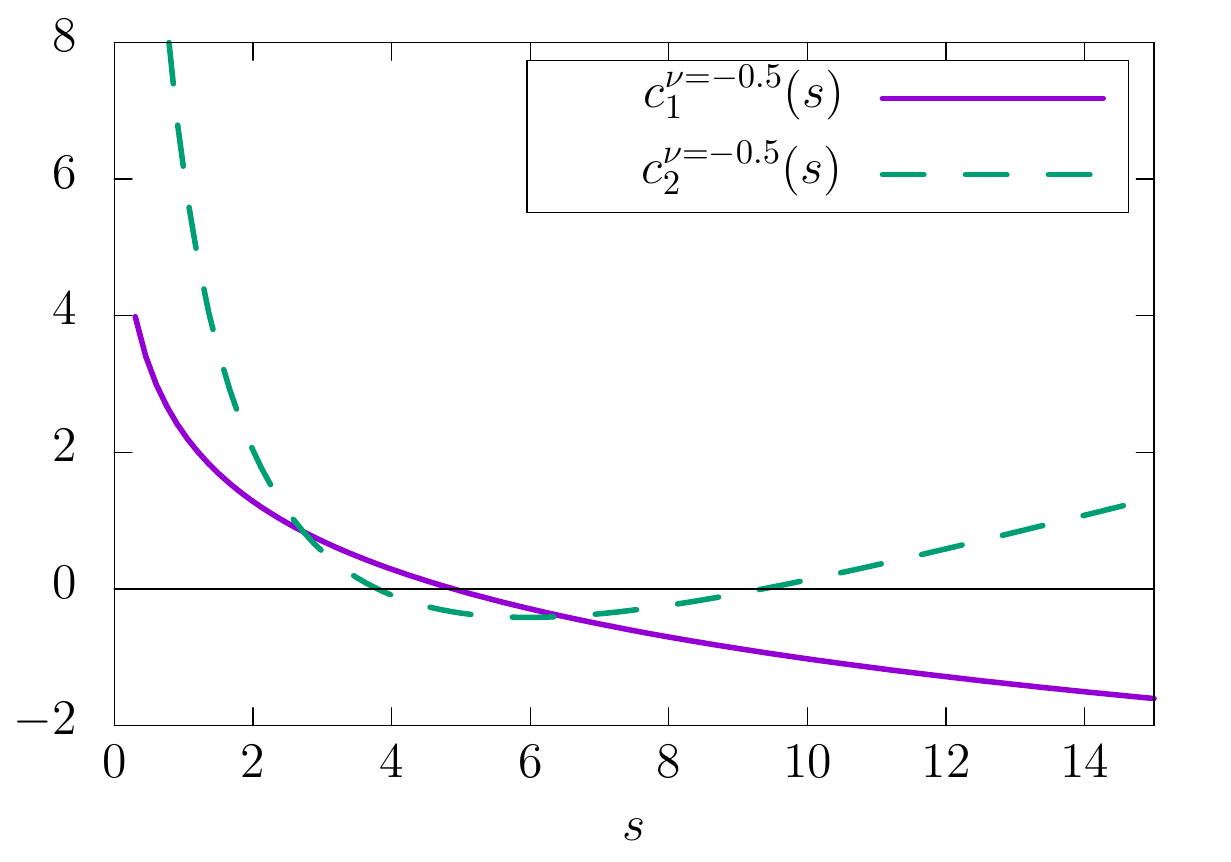}
      \caption{Coefficients $c^\nu_1(s)$ and $c^\nu_2(s)$ for two different
        SF$_\nu$ schemes. Left $\nu=0$, right
        $\nu=-0.5$. The values $s^\star$ defined by the condition
        $c_1(s^\star)=0$ are approximately $s^\star\approx 3$ (left)
        and $s^\star\approx 5$ (right).}
     \label{fig:msbar_nu}
\end{figure}

\begin{figure}
  \centering
    \centering
    \includegraphics[width=0.49\textwidth]{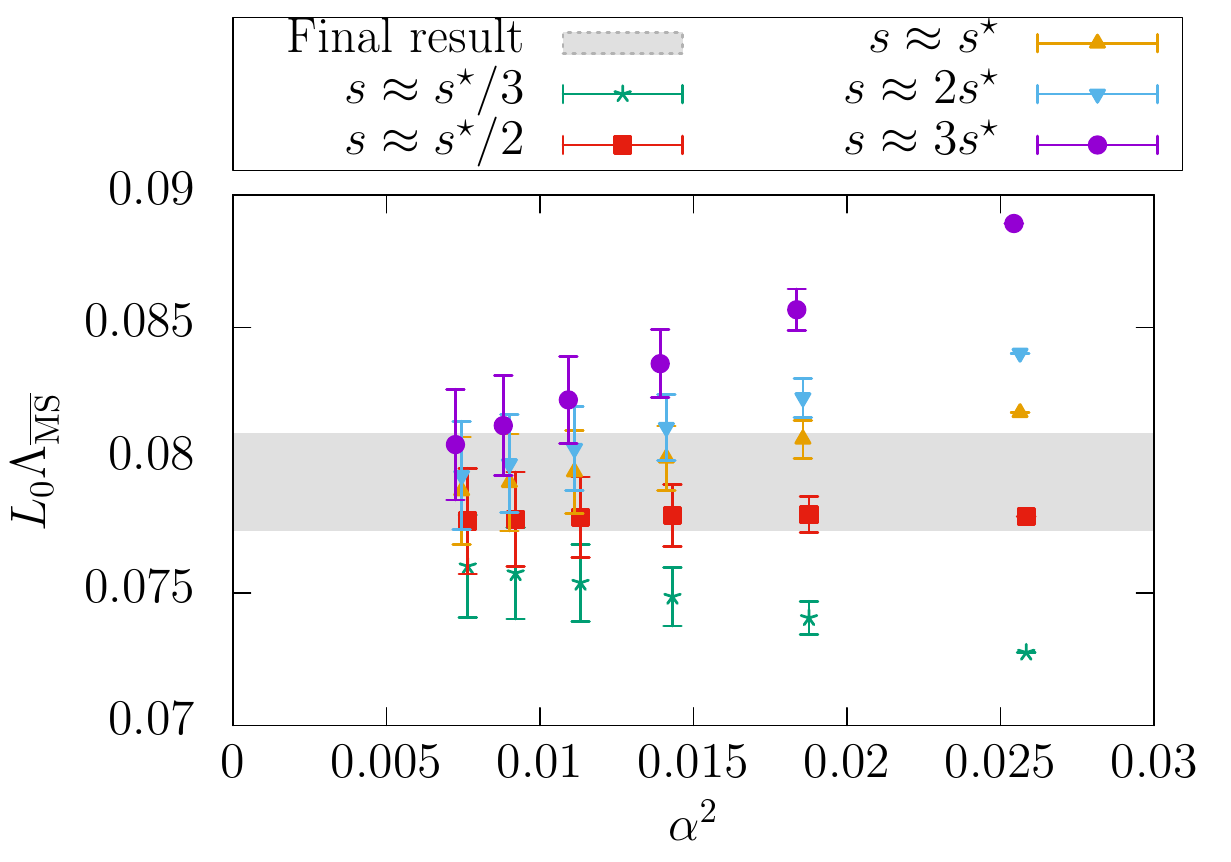}
    \includegraphics[width=0.49\textwidth]{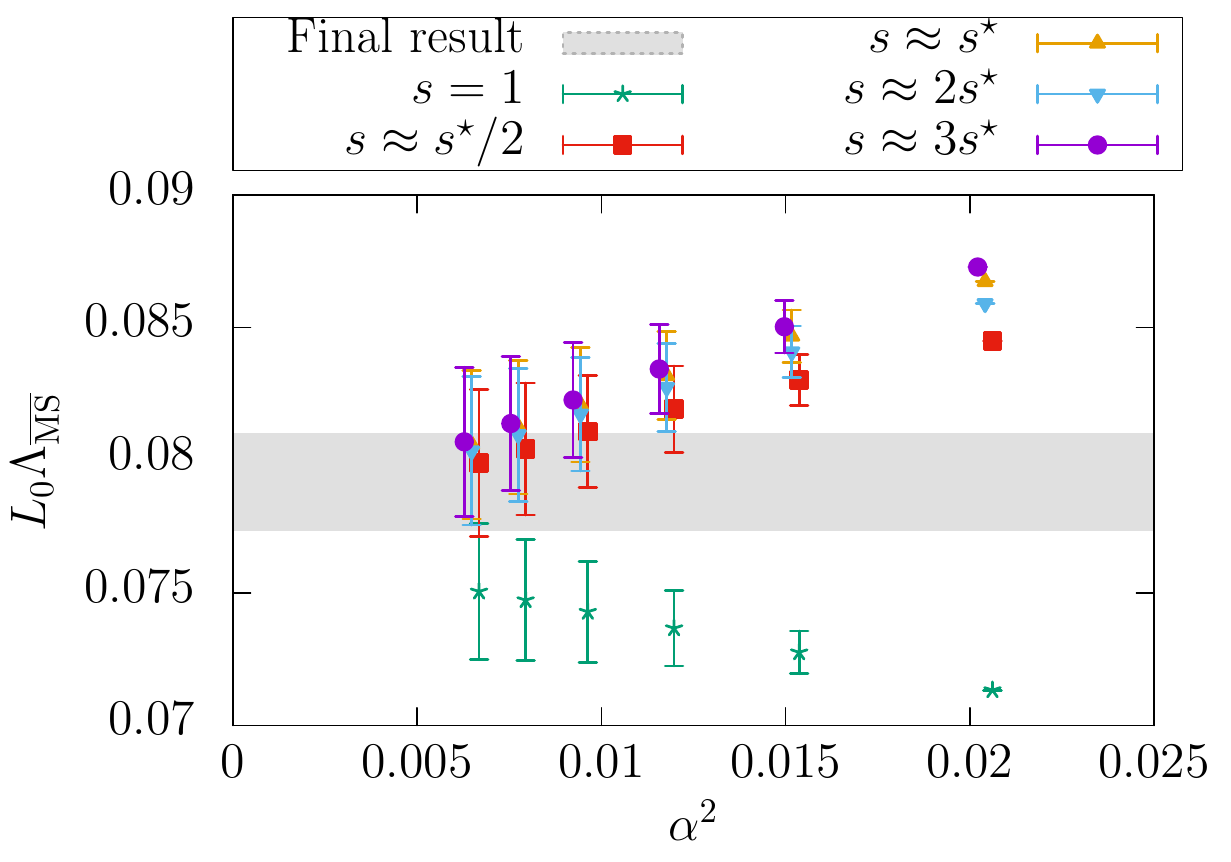}
\caption{Determination of $L_0\Lambda_{\overline{\rm MS}}$ at
  different physical scales (parametrized by the value of $\alpha$ in
  the $x$-axes), and using different renormalization scales (value of
  $s$) to match with the $\MSbar$ scheme. The left (right) panel uses the SF$_\nu$-scheme with 
  $\nu=0$ ($\nu=-0.5$), cf.~text.}
  \label{fig:scale}
\end{figure}

We now use the non-perturbative results for the  SF$_\nu$-couplings
from table~\ref{tab:coupling-Lambda-b3eff} as input in Eq.~(\ref{eq:Lambda-via-MSbar}) 
and study the dependence of the $\Lambda$-parameter estimates 
on the choice of scale $L_n$, the scale factor $s$  and the parameter $\nu$.  
Figure~\ref{fig:scale} shows some typical results;
at fixed $\nu$ and $s$ we observe again an approximate linearity in $\alpha^2$, 
with the asymptotic values being compatible with our best estimate, Eq.~(\ref{eq:final_result}). 
However, we note that the slope varies significantly as a function of $s$ and $\nu$. 

We find that the choice of $s=s^\star$, Eq.~(\ref{eq:scale_ratio}), 
which eliminates the one-loop term in the matching, Eq.~(\ref{eq:FAP}), is often (but not always) 
a good one. For the cases $\nu=0$ and $\nu=-0.5$, figure~\ref{fig:msbar_nu} shows the 1- and 2-loop 
matching coefficients to the $\MSbar$-coupling, Eq.~(\ref{eq:alphaMSbar-alphaSFnu}), as functions
of the scale factor $s$. The values for $s^\star$ are roughly around $3$, $5$ and $2$ for
$\nu=0$, $-0.5$ and $0.3$, respectively. While for $\nu=0$ (similarly for $\nu=0.3$) the
two-loop coefficient is near minimal around $s^\star$ and stays positive 
(figure~\ref{fig:msbar_nu}, left panel), a more complicated pattern is seen for $\nu=-0.5$
(figure~\ref{fig:msbar_nu}, right panel). 

\begin{figure}
  \centering
  \includegraphics[width=\textwidth]{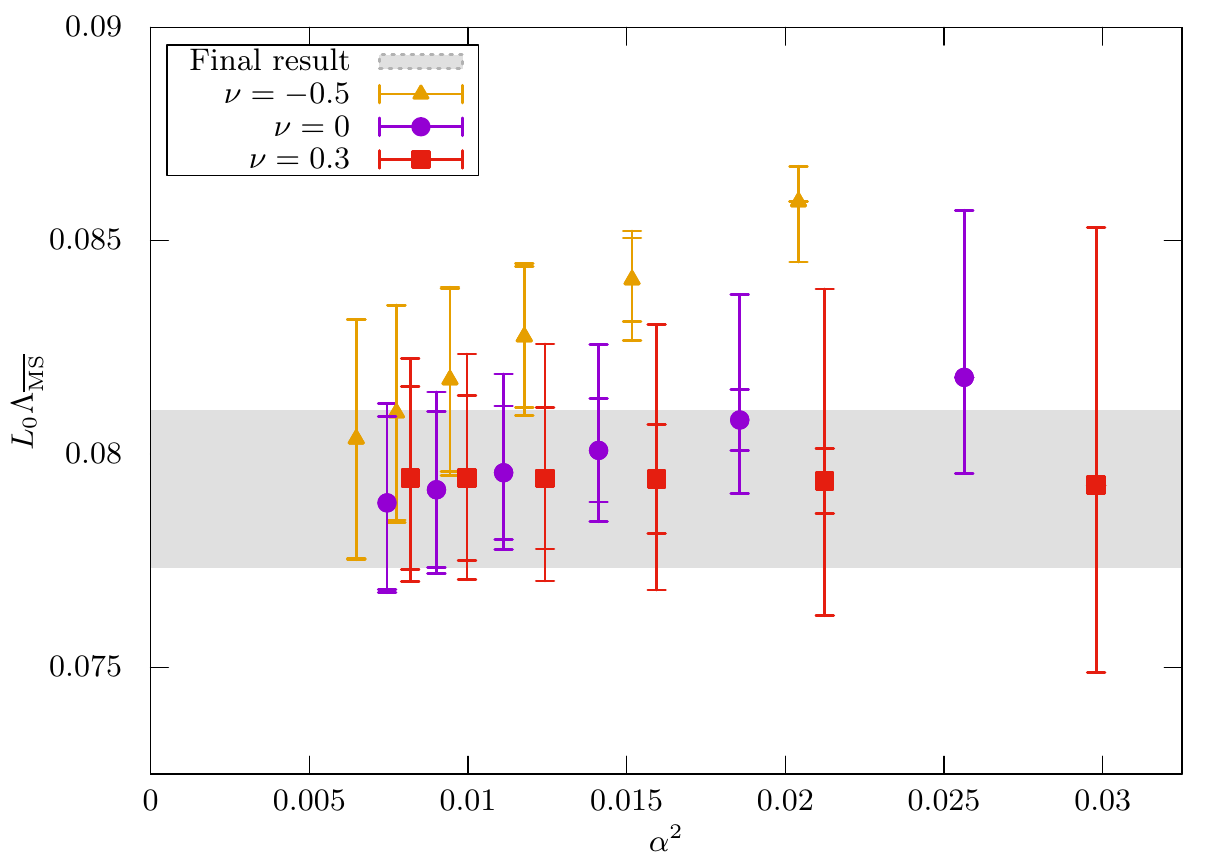}
  \caption{Statistical (interior error band) and total (exterior error
    band) uncertainties in the determination of 
    $L_0\Lambda_{\overline{\rm MS}}$. The total uncertainty is
    computed by adding in quadratures the statistical and systematic
    uncertainties. The latter are computed varying the
    renormalization scale by a factor 2 above and below the value
    $s^\star$. See text for more details.}
  \label{fig:msbar_sys}
\end{figure}

A common method to assign a systematic error to
a perturbative uncertainty consists in a renormalization scale variation
by a factor 2 in both directions, around some ``optimal scale'' (cf.~the review of QCD in \cite{Patrignani:2016xqp}). 
Taking the values $s^\star$ as our optimal values for the scale factor 
we can now assess how this method fares in our context. 
In figure~\ref{fig:msbar_sys} this systematic error is shown,
together with the total errors, for the estimates of the $\Lambda$-parameter. 
As one might expect, this systematic error dominates the error at low energies, 
reduces proportional to $\alpha^2$ and becomes negligible at higher energies.
This is indeed the case for $\nu=0$ and $\nu=0.3$, where the systematic 
errors are seen to bracket the shaded area representing 
our reference result~(\ref{eq:final_result}).
However, this is not the case for $\nu=-0.5$, where
a significant underestimation of the systematic error
is observed, cf.~figure~\ref{fig:msbar_sys}.


However, in all cases we note that for $\alpha\sim 0.1$, the systematic 
uncertainty of the matching with perturbation theory, obtained by varying $s$ is
well below the statistical uncertainties. Moreover the latter are in line with the
errors obtained with our previous strategy. This further reinforces our
previous conclusions: thanks to the high energies reached with the step
scaling method, our uncertainties are fully dominated by statistical
errors, systematic uncertainties being negligible. 
The spread of results obtained by the variation 
of the perturbative matching scale provides  a way to assess the systematic uncertainties which 
works well with the SF schemes at $\nu=0,0.3$, even at $\alpha\approx 0.2$ (although
the systematic uncertainty there is large). But the failure of this method 
for the SF scheme with $\nu=-0.5$ indicates that this method may 
not always be reliable, particularly if the coupling is not small 
and cannot be varied. This illustrates that perturbative truncation 
errors are very difficult to estimate within perturbation theory, 
and that reaching high energies  is crucial for a robust determination of the strong
coupling. Indeed we see that for values $\alpha \approx 0.1$ 
there is nice agreement between all schemes and reasonable 
choices of the scale factor $s$, within errors which clearly allow us
to meet the target accuracy of 3 per cent for the $\Lambda$-parameter.

\section{Conclusions and Outlook}
\label{sec:conclusions}

Using numerical simulations and finite volume step-scaling techniques, 
we have studied a family of SF couplings, parameterized by $\nu$,
over a range of scales corresponding to energies of 4--128 GeV,
thus differing by a scale factor 32. This, together with an unprecedented
control of statistical and systematic errors represents a luxury
which we have exploited to test the accuracy of perturbation theory. Choosing 
the $\Lambda$-parameter for $\nu=0$  in units of $L_0 \approx
1/(4\,\text{GeV})$  as a reference,  
its evaluation requires  the knowledge of a coupling and its $\beta$-function 
between a finite and the infinite energy scale, where the coupling vanishes by
asymptotic freedom. Perturbation theory to 3-loop order is available for the asymptotic
scale dependence beyond an energy scale $1/L$, which can be chosen anywhere in the range
covered by our non-perturbative data, provided the ratio $L/L_0$ is known. By looking
at the spread of values 
for $L_0\Lambda$ one therefore tests the accuracy of perturbation
theory at the scale $1/L$.  
Moreover, the exact relation between $\Lambda$-parameters of different
schemes   requires a one-loop matching of couplings which is known in all cases
considered.  It is therefore also possible to test the robustness of the
$\Lambda$-parameter determination by 
using SF-schemes at various values of $\nu$ as an intermediate step.
The result is neatly illustrated in
Figure~\ref{fig:Lambda-vs-alphasq}, where all data points  
should coincide up to a parametric uncertainty of order $\alpha^2$. We
conclude that 
a target precision of better than $3\%$ for $L_0\Lambda$  (which
approximately corresponds to a $0.5\%$ precision for $\alpha_s(m_Z)$) 
requires non-perturbative data for a large enough range of couplings
so that the perturbative truncation errors can be safely
estimated. Our range of scales $4-128$ GeV reaching down to
$\alpha\approx 0.1$ allows us to reach such a precision.  While some
schemes may give compatible results even at $\alpha \approx 0.2$, it seems all but impossible
to anticipate the quality of a given scheme if the coupling cannot be varied significantly.

With the hindsight of our $2.3\%$ precision result for $L_0\Lambda$, Eq.~(\ref{eq:final_result}), 
we  have also looked at an alternative test, which is close to procedures widely used in
phenomenology. Namely, we have converted our non-perturbative observable, an SF$_\nu$-coupling
with some choice for $\nu$ and $L$, to the $\MSbar$-coupling where we
allowed for a relative scale factor $s$ in this perturbative conversion.
Given the coupling in the $\MSbar$-scheme the full machinery with up to 5-loops for the $\beta$-function~\cite{Chetyrkin:2017bjc,Baikov:2016tgj,Luthe:2017ttc,MS:4loop1,Czakon:2004bu} is available to extract the $\Lambda$-parameter. However, as in phenomenological applications,
the limiting factor is the perturbative order in the conversion to the $\MSbar$-scheme. We can perform
this step at 2-loop order; for comparison we note that the 5-loop, O($\alpha^4$) result
for the $R$-ratio~\cite{Baikov:2012er} translates to 3-loop order when formulated as a conversion between couplings.
Looking at the dependence of the scale factor, a common method consists in identifying an ``optimal scale factor'', $s^\star$, 
and then vary this factor between $s^\star/2$ and $2s^\star$ to obtain a systematic error estimate 
(c.f.~the review of QCD in~\cite{Patrignani:2016xqp}).
It is a bit of an art to determine the ``optimal scale factor'', and some appeal to the kinematics or
the physics of a given observable is often made in this context~\cite{Patrignani:2016xqp}.   
We here applied such a procedure, choosing $s^\star$ close to the ratio of $\Lambda$-parameters, which means the
one-loop coefficient in the conversion to the $\MSbar$ scheme is made very small.
As illustrated in Figure~\ref{fig:msbar_sys}, this procedure gives an error that shrinks proportionally to $\alpha^2$ and 
often brackets the correct result. However, we have also found  cases (e.g.~$\nu=-0.5$) where this
procedure does not work and underestimates the systematic effect
substantially, even at couplings around $\alpha=0.15$.
We interpret this result as a warning: estimating errors within perturbation theory is notoriously difficult,
and one may chance one's luck by being too aggressive in this step.

The work presented in this paper constitutes a major step in the
$\alpha_s$-determination  
by the ALPHA-collaboration~\cite{Bruno:2017gxd}. Despite considerable improvements in the precision, 
this step currently still contributes the largest single error in
this project. One may therefore hope for further progress, perhaps by
combining the SF$_\nu$ schemes with
alternative schemes. Gradient flow couplings are obvious candidates, 
provided the problems with large cutoff effects can be solved~\cite{DallaBrida:2016kgh,Ramos:2015baa}, 
and the perturbative information is pushed at least to the same level as for the SF coupling.
The latter step is possible based on numerical stochastic perturbation theory~~\cite{Brida:2013mva,DallaBrida:2016dai, DallaBrida:2017tru}.
Finally we note that, given the coupling results, similar non-perturbative tests of perturbation theory 
might also be performed using the quark mass parameters~\cite{Campos:2018ahf}.

\addcontentsline{toc}{section}{Acknowledgements}
\begin{acknowledgement}

We thank our colleagues of the ALPHA collaboration, in particular 
C.~Pena, S.~Schaefer, H.~Simma and U.~Wolff for many useful discussions.
We thank the computer centres at HLRN (bep00040) and NIC at DESY, Zeuthen
for providing computing resources and support. 
S.S.~is grateful to the CERN Theory group for the hospitality extended to him. 
P.F.~acknowledges financial support from the Spanish MINECO's ``Centro de
Excelencia Severo Ochoa'' Programme under grant SEV-2012-0249, as well
as from the grant FPA2015-68541-P (MINECO/FEDER). 
This work is based on previous work \cite{Sommer:2015kza} supported strongly by the Deutsche Forschungsgemeinschaft
in the SFB/TR~09. 

\end{acknowledgement}

\newpage
\appendix

\section{Modelling the sensitivity to $\ct$ and $\ctt$}
\label{app:oamodel}
 
In order to estimate the systematic error due to remnant O($a$) effects stemming
from an imperfect tuning of the counterterm coefficients $\ct$ and $\ctt$ 
(cf.~Subsect~\ref{subsec:oamodel}), we model how the data for the step-scaling functions
$\Sigma(u,a/L)$ and  $\Sigma_\nu(u_\nu,a/L)$ were to change
had the simulations been carried out at slightly shifted values of these coefficients.
The basic observables in the simulations are the coupling at $u=\gbar^2(L)$ and $\vbar$ 
from which the  couplings at $\nu\ne 0$ and the step-scaling functions are constructed. 
We therefore first obtain a model for the $\ct$- and $\ctt$-dependence of 
$u$ and $\vbar$. In a second step we translate this information to $u_\nu$ and finally
to the step-scaling functions.

\subsection{Linear $a$-effects in $\gbar^2$ and $\vbar$}

\begin{figure}[t]
   \centering
   \includegraphics[height=0.45\textwidth]{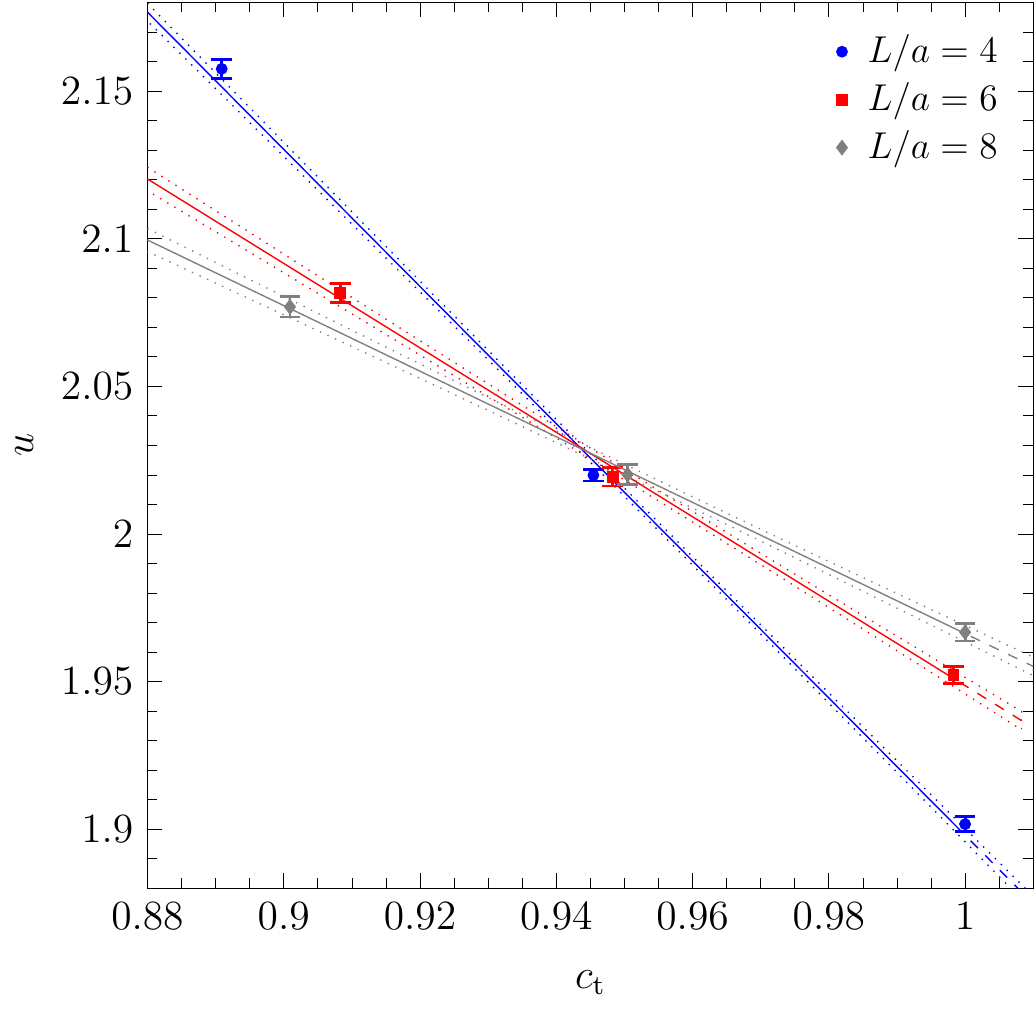}\quad
   \includegraphics[height=0.45\textwidth]{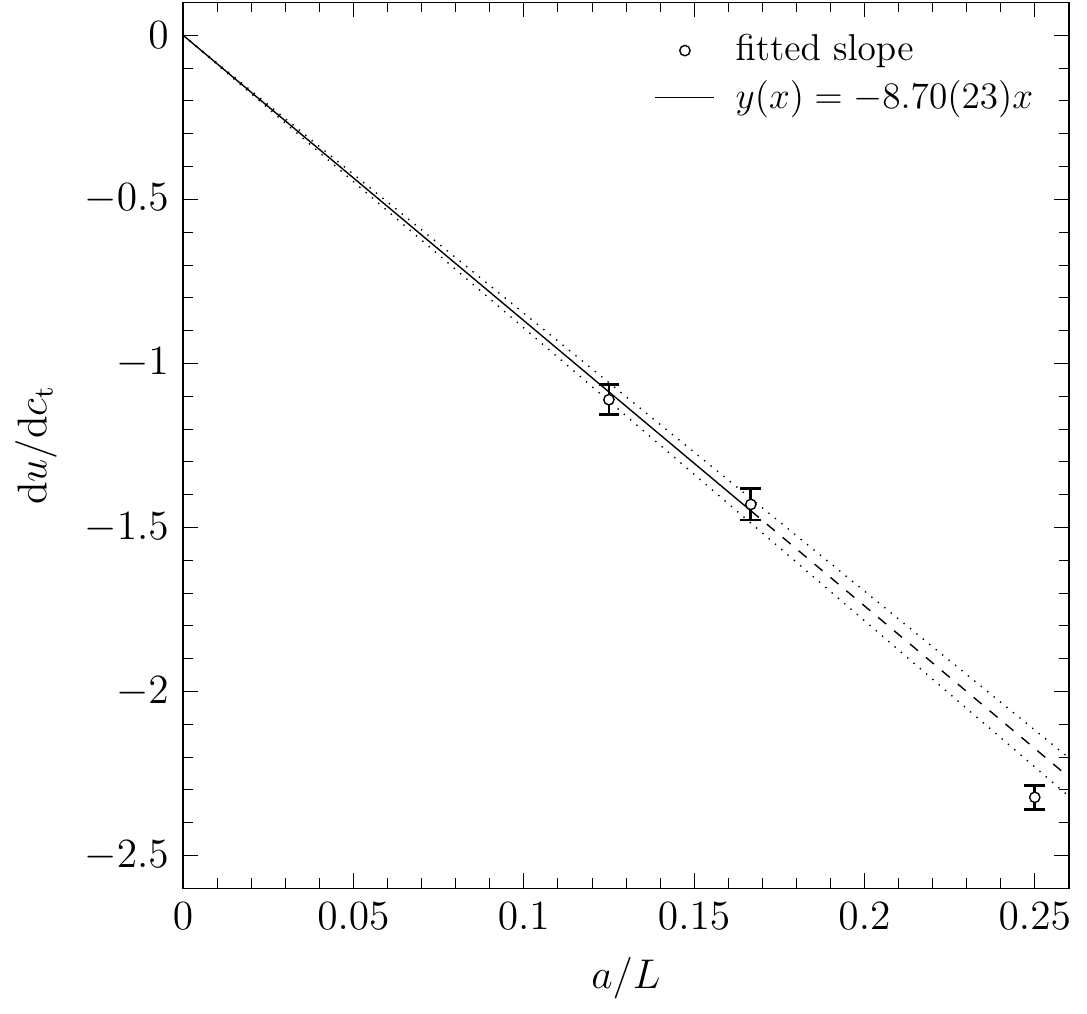}
   \caption{The left panel shows the dependence on $\ct$ of the coupling
            $u=\gbar^2(L)$. The improvement coefficient is varied around its
            2-loop value and the value of the coupling is $u\approx 2$. The
            right panel shows the slopes vs.~$a/L$.
           }
   \label{fig:dudct}
\end{figure}

We have carried out simulations at fixed $u=2.02$, lattice sizes $L/a=4,6,8$ and
with $\ct$- and $\ctt$-values varied around their perturbative default values.
For instance, fig.~\ref{fig:dudct} shows the data for the $\ct$-dependence of $u=\gbar^2(L)$.
These results are then used for numerical estimates of the respective derivatives 
which are collected in table~\ref{tab:ctctt-derivs-MC}. 
\begin{table}[t]
  \centering
  \small
  \begin{tabular}[c]{ccccc}\toprule
    $L/a$   & $\dd u/ \dd\ct$ & $\dd u/\dd\ctt$ & $\dd \vbar /\dd\ct$ &
    $\dd \vbar /\dd \ctt$ \\\midrule
      4 &$-2.323(37)$ & 0.5523(20) & 0.036(19) & $-0.147(11)$ \\     
      6 &$-1.430(48)$ & 0.4771(22) & 0.045(24) & $-0.089(11)$ \\ 
      8 &$-1.111(46)$ & 0.5368(31) & 0.017(23) & $-0.096(15)$ \\
    \bottomrule
  \end{tabular}
  \caption{Non-perturbative estimates for the derivatives with respect to $\ct$
           and $\ctt$ at $u= 2.02$. The derivatives were obtained by numerical
           variation of the improvement coefficients around their perturbative
           2-loop and 1-loop values, respectively.
          }
  \label{tab:ctctt-derivs-MC}
\end{table}
Defining the functions
\begin{equation}
        \delta_b(u) = \frac{L}{2a} \frac{1}{u} \frac{\dd u}{\dd \ct} \,, \qquad  
   \varepsilon_b(u) = \frac{L}{a}\left.\frac{\dd \vbar}{\dd \ct}\right\vert_{\gbar^2(L)=u} \,,  
\end{equation}
and analogous functions $\tilde\delta_b(u)$ and $\tilde\varepsilon_b(u)$, with the derivative taken with respect to
$\ctt$ instead of $\ct$. From the simulations results one then infers their values at $u=2.02$,
\begin{equation}
   \delta_b(2.02)= -2.15(6)\,,\qquad
   \varepsilon_b(2.02) = 0.22(11) \,,
\end{equation}
and, for the $\ctt$-dependence,
\begin{equation}
   \tilde\delta_b(2.02) = 0.785(4)\,,\qquad
   \tilde\varepsilon_b(2.02) = -0.59(6)\,.
\end{equation}
In order to obtain these functions for a range of $u$-values we interpolate these non-perturbative results
with perturbation theory up to 2-loop order \cite{Bode:1999sm} and arrive at 
\begin{eqnarray}
   \delta_b(u) &=& -\left(1 + 0.40 u + 0.085(14) u^2\right)\,, \label{eq:delb}\\
   \varepsilon_b(u) &=& 0.11(6)u\,, \label{eq:varepsb}\\
   \tilde\delta_b(u) &=&  0.25 u + 0.0686(7)u^2\,,\label{eq:tdelb}\\
   \tilde\varepsilon_b(u) &=& -1.35+0.38(3)u\,.  \label{eq:tvarepsb}
\end{eqnarray}
We remark that the separation of the $a/L$- and the $u$-dependence is
only approximate; in perturbation theory this neglects any logarithmic dependence on $L/a$
in the coefficients. Furthermore, while we have indicated the error in the fit coefficients, 
the functions will be used with the central values throughout and these errors will not be propagated.
We observe that the dominant effect for the coupling is the $\ct$-dependence. In perturbation theory this
is reflected by the fact that the fermionic $\ctt$-counterterm only contributes via loop effects
and is thus suppressed by a relative factor $u$. For $\vbar$ the hierarchy is reversed, 
due to the fact that the $\ct$-counterterm is $\nu$-independent
and does not contribute to $v_1$, in contrast to the $\ctt$-counterterm.

With these definitions, we may now generalize the function $\delta_b(u)$
to the SF$_\nu$ coupling based on the relation $u_\nu = u/(1-\nu u \vbar)$,
\begin{equation}
  \delta_{b,\nu}(u_\nu) = u_\nu\left(\frac{\delta_b(u)}{u} + \frac12 \nu\varepsilon_b(u)\right) \,,
  \label{eq:delbdelbnu}
\end{equation}
and the function  $\tilde\delta_{b,\nu}$ is defined analogously.
Here, explicit values for  $\delta_{b,\nu}(u_\nu)$ require 
both $u$ and $u_\nu$, which could be obtained from the data via $\beta$
and $L/a$. However, we prefer to define explicit functions of $u_\nu$ 
which is obtained by inverting numerically the  function
\begin{equation}
  u_\nu(u) = \frac{u}{1-\nu u \omega(u)}\,,\quad \Rightarrow \quad u = u(u_\nu)\,,
\end{equation}
with $\omega(u)$ from Eq.~(\ref{eq:omegafit}), and by then
substituting for $u$ in Eq.~(\ref{eq:delbdelbnu}). Obviously, this model for the
$\ct$- and $\ctt$-derivatives neglects some cutoff 
effects in the derivatives which are of higher than linear order in $a/L$.

\subsection{Step-scaling functions}

Given the models for the couplings $\gbar^2(L)$ we now translate this
information to the step-scaling function, which is defined as the coupling $u'=\gbar^2(2L)$, taken
as a function of $u=\gbar^2(L)$. Neglecting higher order cutoff effects we then find 
\begin{equation}
  \left.\frac{\dd \Sigma(u,a/L)}{\dd\ct}\right\vert_{u=\gbar^2(L)} 
   \approx \frac{a}{L}\left(\sigma(u)\delta_b(\sigma(u)) - 2u\sigma'(u)\delta_b(u)\right)\,, 
\end{equation}
and analogous equations hold for $\nu\ne 0$ and the derivatives with respect to $\ctt$, with
the obvious substitutions. We furthermore specify that we insert
the continuum step-scaling functions to their known perturbative order,
\begin{equation}
  \sigma(u) = u + s_0 u^2 + s_1 u^3 + s_2 u^4
\end{equation}
with $s_{0,1,2}$ as given in Eq.~(\ref{eq:s012}).
With this convention we then define the sensitivity in Eq.~(\ref{eq:Sigmact}) as
\begin{eqnarray}
  \delta_{\ct}\Sigma_\nu(u_\nu,a/L) &=& \frac{a}{L}\left(\sigma_\nu(u_\nu) \delta_{b,\nu}(\sigma_\nu(u_\nu)) - 
   2u_\nu\sigma'_\nu(u_\nu)\delta_{b,\nu}(u_\nu)\right)\,,  
\end{eqnarray}
and analogously for $\delta_{\ctt}\Sigma_\nu$, with the replacements 
$\delta_b \rightarrow \tilde\delta_b$.
The case $\nu=0$ is included by omitting the index $\nu$. Furthermore, we define
\begin{equation}
  \delta_{\ct}\Omega(u,a/L) = \delta_{\ct}\tilde \Omega (u,a/L) = \frac{a}{L}\varepsilon_{b}(u),
\end{equation}
and analogously for $\delta_{\ctt}\Omega = \delta_{\ctt}\tilde\Omega$ with the replacement 
$\varepsilon_b \rightarrow \tilde\varepsilon_b$.
The functions $\varepsilon_b$ and $\tilde{\varepsilon}_b$ are given in Eqs.~(\ref{eq:varepsb},\ref{eq:tvarepsb}).
The data $\tilde{\Omega}(u,a/L)$, obtained from the $2L$-lattices, are treated 
exactly like $\Omega(u,a/L)$, as both the $\ct$- and $\ctt$-variations were found to be
rather insensitive to the precise definition of the critical mass 
(at least in perturbation theory).

We finally note that a slight extrapolation of our functions $\delta_b,\tilde\delta_b$ 
and $\varepsilon_b,\tilde\varepsilon_b$ beyond $u\approx 2$ is implicit due to the largest argument being 
$\sigma(2)\approx 2.45$. However, we expect that the model is
not too far off, and are here not overly concerned about errors at the 10 percent level, which
would add a corresponding uncertainty on our systematic error estimates.
If future precision studies require a more refined model for these systematic effects, 
one might revisit the separation of the variables $L/a$ and $u$ in our models, and also check the non-perturbative 
derivatives at the largest relevant arguments. Finally we remark that this whole exercise would become 
redundant if the counterterm coefficients $\ct$ and $\ctt$ could be determined non-perturbatively
as functions of $g_0$.

\usebiblio{SF-paper}{}


\providecommand{\href}[2]{#2}\begingroup\raggedright\begin{thebibliography}{10}

\bibitem{Brida:2016flw}
\textsc{ALPHA collaboration}, M.~Dalla~Brida, P.~Fritzsch, T.~Korzec, A.~Ramos,
  S.~Sint, and R.~Sommer, {\it {Determination of the QCD $\Lambda$-parameter
  and the accuracy of perturbation theory at high energies}},  {\em Phys. Rev.
  Lett.} {\bf 117} (2016), no.~18 182001,
  [\href{http://arxiv.org/abs/1604.06193}{{\tt 1604.06193}}].

\bibitem{Patrignani:2016xqp}
\textsc{Particle Data Group collaboration}, C.~Patrignani et~al., {\it {Review
  of Particle Physics}},  {\em Chin. Phys.} {\bf C40} (2016), no.~10 100001.

\bibitem{Chetyrkin:2017bjc}
K.~G. Chetyrkin, G.~Falcioni, F.~Herzog, and J.~A.~M. Vermaseren, {\it
  {Five-loop renormalisation of QCD in covariant gauges}},  {\em JHEP} {\bf 10}
  (2017) 179, [\href{http://arxiv.org/abs/1709.08541}{{\tt 1709.08541}}].
  [Addendum: JHEP 12 (2017) 006].

\bibitem{Baikov:2016tgj}
P.~A. Baikov, K.~G. Chetyrkin, and J.~H. K{\"u}hn, {\it {Five-Loop Running of
  the QCD coupling constant}},  {\em Phys. Rev. Lett.} {\bf 118} (2017), no.~8
  082002, [\href{http://arxiv.org/abs/1606.08659}{{\tt 1606.08659}}].

\bibitem{Luthe:2017ttc}
T.~Luthe, A.~Maier, P.~Marquard, and Y.~Schr{\"o}der, {\it {Complete
  renormalization of QCD at five loops}},  {\em JHEP} {\bf 03} (2017) 020,
  [\href{http://arxiv.org/abs/1701.07068}{{\tt 1701.07068}}].

\bibitem{MS:4loop1}
T.~van Ritbergen, J.~A.~M. Vermaseren, and S.~A. Larin, {\it The four loop beta
  function in quantum chromodynamics},  {\em Phys. Lett.} {\bf B400} (1997)
  379--384, [\href{http://arxiv.org/abs/hep-ph/9701390}{{\tt hep-ph/9701390}}].

\bibitem{Czakon:2004bu}
M.~Czakon, {\it {The Four-loop QCD beta-function and anomalous dimensions}},
  {\em Nucl.Phys.} {\bf B710} (2005) 485--498,
  [\href{http://arxiv.org/abs/hep-ph/0411261}{{\tt hep-ph/0411261}}].

\bibitem{Salam:2017qdl}
G.~P. Salam, {\it {The strong coupling: a theoretical perspective}},
  \href{http://arxiv.org/abs/1712.05165}{{\tt 1712.05165}}.

\bibitem{Luscher:1985dn}
M.~L{\"u}scher, {\it {Volume Dependence of the Energy Spectrum in Massive
  Quantum Field Theories. 1. Stable Particle States}},  {\em Commun. Math.
  Phys.} {\bf 104} (1986) 177.

\bibitem{Luscher:1991wu}
M.~L{\"u}scher, P.~Weisz, and U.~Wolff, {\it {A numerical method to compute the
  running coupling in asymptotically free theories}},  {\em Nucl. Phys.} {\bf
  B359} (1991) 221--243.

\bibitem{Jansen:1995ck}
K.~Jansen, C.~Liu, M.~L{\"u}scher, H.~Simma, S.~Sint, R.~Sommer, P.~Weisz, and
  U.~Wolff, {\it {Nonperturbative renormalization of lattice QCD at all
  scales}},  {\em Phys. Lett.} {\bf B372} (1996) 275--282,
  [\href{http://arxiv.org/abs/hep-lat/9512009}{{\tt hep-lat/9512009}}].

\bibitem{Bruno:2017gxd}
\textsc{ALPHA collaboration}, M.~Bruno, M.~Dalla~Brida, P.~Fritzsch, T.~Korzec,
  A.~Ramos, S.~Schaefer, H.~Simma, S.~Sint, and R.~Sommer, {\it {QCD Coupling
  from a Nonperturbative Determination of the Three-Flavor $\Lambda$
  Parameter}},  {\em Phys. Rev. Lett.} {\bf 119} (2017), no.~10 102001,
  [\href{http://arxiv.org/abs/1706.03821}{{\tt 1706.03821}}].

\bibitem{DallaBrida:2016kgh}
\textsc{ALPHA collaboration}, M.~Dalla~Brida, P.~Fritzsch, T.~Korzec, A.~Ramos,
  S.~Sint, and R.~Sommer, {\it {Slow running of the Gradient Flow coupling from
  200 MeV to 4 GeV in $N_{\rm f}=3$ QCD}},  {\em Phys. Rev.} {\bf D95} (2017),
  no.~1 014507, [\href{http://arxiv.org/abs/1607.06423}{{\tt 1607.06423}}].

\bibitem{Bode:1998hd}
\textsc{ALPHA collaboration}, A.~Bode, U.~Wolff, and P.~Weisz, {\it {Two loop
  computation of the Schr{\"o}dinger functional in pure SU(3) lattice gauge
  theory}},  {\em Nucl. Phys.} {\bf B540} (1999) 491--499,
  [\href{http://arxiv.org/abs/hep-lat/9809175}{{\tt hep-lat/9809175}}].

\bibitem{Bode:1999sm}
\textsc{ALPHA collaboration}, A.~Bode, P.~Weisz, and U.~Wolff, {\it {Two loop
  computation of the Schr{\"o}dinger functional in lattice QCD}},  {\em Nucl.
  Phys.} {\bf B576} (2000) 517--539,
  [\href{http://arxiv.org/abs/hep-lat/9911018}{{\tt hep-lat/9911018}}].
  [Erratum: Nucl. Phys. B608 (2001) 481].

\bibitem{Christou:1998wk}
C.~Christou, H.~Panagopoulos, A.~Feo, and E.~Vicari, {\it {The two loop
  relation between the bare lattice coupling and the ${\overline{\rm MS}}$
  coupling in QCD with Wilson fermions}},  {\em Phys. Lett.} {\bf B426} (1998)
  121--124.

\bibitem{Christou:1998ws}
C.~Christou, A.~Feo, H.~Panagopoulos, and E.~Vicari, {\it {The three loop
  $\beta$-function of $SU(N)$ lattice gauge theories with Wilson fermions}},
  {\em Nucl. Phys.} {\bf B525} (1998) 387--400,
  [\href{http://arxiv.org/abs/hep-lat/9801007}{{\tt hep-lat/9801007}}].
  [Erratum: Nucl. Phys. B608 (2001) 479].

\bibitem{Capitani:1998mq}
\textsc{ALPHA collaboration}, S.~Capitani, M.~{L\"uscher}, R.~Sommer, and
  H.~Wittig, {\it {Nonperturbative quark mass renormalization in quenched
  lattice QCD}},  {\em Nucl. Phys.} {\bf B544} (1999) 669--698,
  [\href{http://arxiv.org/abs/hep-lat/9810063}{{\tt hep-lat/9810063}}].

\bibitem{DellaMorte:2004bc}
\textsc{ALPHA collaboration}, M.~Della~Morte, R.~Frezzotti, J.~Heitger,
  J.~Rolf, R.~Sommer, and U.~Wolff, {\it {Computation of the strong coupling in
  QCD with two dynamical flavors}},  {\em Nucl. Phys.} {\bf B713} (2005)
  378--406, [\href{http://arxiv.org/abs/hep-lat/0411025}{{\tt
  hep-lat/0411025}}].

\bibitem{Aoki:2009tf}
\textsc{PACS-CS collaboration}, S.~Aoki et~al., {\it {Precise determination of
  the strong coupling constant in $N_f = 2+1$ lattice QCD with the
  Schr\"odinger functional scheme}},  {\em JHEP} {\bf 0910} (2009) 053,
  [\href{http://arxiv.org/abs/0906.3906}{{\tt 0906.3906}}].

\bibitem{Tekin:2010mm}
\textsc{ALPHA collaboration}, F.~Tekin, R.~Sommer, and U.~Wolff, {\it {The
  Running coupling of QCD with four flavors}},  {\em Nucl. Phys.} {\bf B840}
  (2010) 114--128, [\href{http://arxiv.org/abs/1006.0672}{{\tt 1006.0672}}].

\bibitem{Weinberg:1951ss}
S.~Weinberg, {\it {New approach to the renormalization group}},  {\em Phys.
  Rev.} {\bf D8} (1973) 3497--3509.

\bibitem{Luscher:1992an}
M.~L{\"u}scher, R.~Narayanan, P.~Weisz, and U.~Wolff, {\it {The Schr{\"o}dinger
  functional: a renormalizable probe for non-Abelian gauge theories}},  {\em
  Nucl. Phys.} {\bf B384} (1992) 168--228,
  [\href{http://arxiv.org/abs/hep-lat/9207009}{{\tt hep-lat/9207009}}].

\bibitem{Sint:1993un}
S.~Sint, {\it {On the Schr{\"o}dinger functional in QCD}},  {\em Nucl. Phys.}
  {\bf B421} (1994) 135--158, [\href{http://arxiv.org/abs/hep-lat/9312079}{{\tt
  hep-lat/9312079}}].

\bibitem{Luscher:1993gh}
M.~L{\"u}scher, R.~Sommer, P.~Weisz, and U.~Wolff, {\it {A Precise
  determination of the running coupling in the SU(3) Yang-Mills theory}},  {\em
  Nucl. Phys.} {\bf B413} (1994) 481--502,
  [\href{http://arxiv.org/abs/hep-lat/9309005}{{\tt hep-lat/9309005}}].

\bibitem{Sint:1995ch}
S.~Sint and R.~Sommer, {\it {The running coupling from the QCD Schr{\"o}dinger
  functional: a one loop analysis}},  {\em Nucl. Phys.} {\bf B465} (1996)
  71--98, [\href{http://arxiv.org/abs/hep-lat/9508012}{{\tt hep-lat/9508012}}].

\bibitem{Sint:2012ae}
S.~Sint and P.~Vilaseca, {\it {Lattice artefacts in the Schr{\"o}dinger
  functional coupling for strongly interacting theories}},  {\em PoS} {\bf
  LATTICE2012} (2012) 031, [\href{http://arxiv.org/abs/1211.0411}{{\tt
  1211.0411}}].

\bibitem{Luscher:2010iy}
M.~L{\"u}scher, {\it {Properties and uses of the Wilson flow in lattice QCD}},
  {\em JHEP} {\bf 08} (2010) 071, [\href{http://arxiv.org/abs/1006.4518}{{\tt
  1006.4518}}]. [Erratum: JHEP 03 (2014) 092].

\bibitem{Fritzsch:2013je}
P.~Fritzsch and A.~Ramos, {\it {The gradient flow coupling in the
  Schr{\"o}dinger functional}},  {\em JHEP} {\bf 10} (2013) 008,
  [\href{http://arxiv.org/abs/1301.4388}{{\tt 1301.4388}}].

\bibitem{Harlander:2016vzb}
R.~V. Harlander and T.~Neumann, {\it {The perturbative QCD gradient flow to
  three loops}},  {\em JHEP} {\bf 06} (2016) 161,
  [\href{http://arxiv.org/abs/1606.03756}{{\tt 1606.03756}}].

\bibitem{Brida:2013mva}
M.~Dalla~Brida and D.~Hesse, {\it {Numerical Stochastic Perturbation Theory and
  the Gradient Flow}},  {\em PoS} {\bf Lattice2013} (2014) 326,
  [\href{http://arxiv.org/abs/1311.3936}{{\tt 1311.3936}}].

\bibitem{DallaBrida:2016dai}
M.~Dalla~Brida and M.~L{\"u}scher, {\it {The gradient flow coupling from
  numerical stochastic perturbation theory}},  {\em PoS} {\bf LATTICE2016}
  (2016) 332, [\href{http://arxiv.org/abs/1612.04955}{{\tt 1612.04955}}].

\bibitem{DallaBrida:2017tru}
M.~Dalla~Brida and M.~L{\"u}scher, {\it {SMD-based numerical stochastic
  perturbation theory}},  {\em Eur. Phys. J.} {\bf C77} (2017), no.~5 308,
  [\href{http://arxiv.org/abs/1703.04396}{{\tt 1703.04396}}].

\bibitem{Beneke:1998ui}
M.~Beneke, {\it {Renormalons}},  {\em Phys. Rept.} {\bf 317} (1999) 1--142,
  [\href{http://arxiv.org/abs/hep-ph/9807443}{{\tt hep-ph/9807443}}].

\bibitem{Sheikholeslami:1985ij}
B.~Sheikholeslami and R.~Wohlert, {\it {Improved Continuum Limit Lattice Action
  for QCD with Wilson Fermions}},  {\em Nucl. Phys.} {\bf B259} (1985) 572.

\bibitem{Luscher:1996sc}
M.~{L\"uscher}, S.~Sint, R.~Sommer, and P.~Weisz, {\it {Chiral symmetry and
  O(a) improvement in lattice QCD}},  {\em Nucl. Phys.} {\bf B478} (1996)
  365--400, [\href{http://arxiv.org/abs/hep-lat/9605038}{{\tt
  hep-lat/9605038}}].

\bibitem{Yamada:2004ja}
\textsc{JLQCD, CP-PACS collaboration}, N.~Yamada et~al., {\it {Non-perturbative
  O($a$)-improvement of Wilson quark action in three-flavor QCD with plaquette
  gauge action}},  {\em Phys. Rev.} {\bf D71} (2005) 054505,
  [\href{http://arxiv.org/abs/hep-lat/0406028}{{\tt hep-lat/0406028}}].

\bibitem{Luscher:1996ug}
M.~L{\"u}scher, S.~Sint, R.~Sommer, P.~Weisz, and U.~Wolff, {\it
  {Nonperturbative O(a) improvement of lattice QCD}},  {\em Nucl. Phys.} {\bf
  B491} (1997) 323--343, [\href{http://arxiv.org/abs/hep-lat/9609035}{{\tt
  hep-lat/9609035}}].

\bibitem{Luscher:1996vw}
M.~{L\"uscher} and P.~Weisz, {\it {O(a) improvement of the axial current in
  lattice QCD to one loop order of perturbation theory}},  {\em Nucl. Phys.}
  {\bf B479} (1996) 429--458, [\href{http://arxiv.org/abs/hep-lat/9606016}{{\tt
  hep-lat/9606016}}].

\bibitem{Sint:1997jx}
S.~Sint and P.~Weisz, {\it {Further results on O(a) improved lattice QCD to one
  loop order of perturbation theory}},  {\em Nucl. Phys.} {\bf B502} (1997)
  251--268, [\href{http://arxiv.org/abs/hep-lat/9704001}{{\tt
  hep-lat/9704001}}].

\bibitem{deDivitiis:1994yz}
\textsc{ALPHA collaboration}, G.~de~Divitiis, R.~Frezzotti, M.~Guagnelli,
  M.~{L\"uscher}, R.~Petronzio, R.~Sommer, P.~Weisz, and U.~Wolff, {\it
  {Universality and the approach to the continuum limit in lattice gauge
  theory}},  {\em Nucl. Phys.} {\bf B437} (1995) 447--470,
  [\href{http://arxiv.org/abs/hep-lat/9411017}{{\tt hep-lat/9411017}}].

\bibitem{Luscher:2012av}
M.~{L\"uscher} and S.~Schaefer, {\it {Lattice QCD with open boundary conditions
  and twisted-mass reweighting}},  {\em Comput. Phys. Commun.} {\bf 184} (2013)
  519--528, [\href{http://arxiv.org/abs/1206.2809}{{\tt 1206.2809}}].

\bibitem{openqcd:2013}
open{QCD}: Simulation program for~lattice {QCD}.
  \texttt{http://luscher.web.cern.ch/luscher/openQCD/}.

\bibitem{Wolff:2003sm}
\textsc{ALPHA collaboration}, U.~Wolff, {\it {Monte Carlo errors with less
  errors}},  {\em Comput. Phys. Commun.} {\bf 156} (2004) 143--153,
  [\href{http://arxiv.org/abs/hep-lat/0306017}{{\tt hep-lat/0306017}}].
  [Erratum: Comput. Phys. Commun. 176 (2007) 383].

\bibitem{fritzsch:2018}
P.~Fritzsch and T.~Korzec, {\it {Simulating the QCD Schr{\"o}dinger functional
  with three massless quark flavors}},  {\em {\emph{in preparation}}} (2018).

\bibitem{Baikov:2012er}
P.~A. Baikov, K.~G. Chetyrkin, J.~H. Kuhn, and J.~Rittinger, {\it {Complete
  ${\cal O}(\alpha_s^4)$ QCD Corrections to Hadronic $Z$-Decays}},  {\em Phys.
  Rev. Lett.} {\bf 108} (2012) 222003,
  [\href{http://arxiv.org/abs/1201.5804}{{\tt 1201.5804}}].

\bibitem{Ramos:2015baa}
A.~Ramos and S.~Sint, {\it {Symanzik improvement of the gradient flow in
  lattice gauge theories}},  {\em Eur. Phys. J.} {\bf C76} (2016), no.~1 15.

\bibitem{Campos:2018ahf}
\textsc{ALPHA collaboration}, I.~Campos, P.~Fritzsch, C.~Pena, D.~Preti,
  A.~Ramos, and A.~Vladikas, {\it {Non-perturbative quark mass renormalisation
  and running in $N_f=3$ QCD}},  \href{http://arxiv.org/abs/1802.05243}{{\tt
  1802.05243}}.

\bibitem{Sommer:2015kza}
R.~Sommer and U.~Wolff, {\it {Non-perturbative computation of the strong
  coupling constant on the lattice}},  {\em Nucl. Part. Phys. Proc.} {\bf
  261-262} (2015) 155--184, [\href{http://arxiv.org/abs/1501.01861}{{\tt
  1501.01861}}].

\end{thebibliography}\endgroup
\end{document}